\documentclass[aps,final,letterpaper,10pt,twocolumn,floats,showpacs,amsmath,amsfonts,amssymb]{revtex4-1}

\usepackage{graphicx}
\usepackage[colorlinks=true,citecolor=blue]{hyperref}
\usepackage[caption=false]{subfig}
\usepackage{pstricks}
\usepackage{amsmath}
\usepackage{amsbsy}
\begin{document}

\preprint{APS/123-QED}

\title{Waiting time distribution revealing the internal spin dynamics in a double quantum dot}

\author{Krzysztof Ptaszy\'{n}ski}
 \email{krzysztof.ptaszynski@ifmpan.poznan.pl}
\affiliation{%
 Institute of Molecular Physics, Polish Academy of Sciences, ul. M. Smoluchowskiego 17, 60-179 Pozna\'{n}, Poland
}%

\date{\today}

\begin{abstract}
Waiting time distribution and the zero-frequency full counting statistics of unidirectional electron transport through a double quantum dot molecule attached to spin-polarized leads are analyzed using the quantum master equation. The waiting time distribution exhibits a non-trivial dependence on the value of the exchange coupling between the dots and the gradient of the applied magnetic field, which reveals the oscillations between the spin states of the molecule. The zero-frequency full counting statistics, on the other hand, is independent of the aforementioned quantities, thus giving no insight into the internal dynamics. The fact that the waiting time distribution and the zero-frequency full counting statistics give a non-equivalent information is associated with two factors. Firstly, it can be explained by the sensitivity to different timescales of the dynamics of the system. Secondly, it is associated with the presence of the correlation between subsequent waiting times, which makes the renewal theory, relating the full counting statistics and the waiting time distribution, not longer applicable. The study highlights the particular usefulness of the waiting time distribution for the analysis of the internal dynamics of mesoscopic systems.

\end{abstract}


\maketitle


\section{\label{sec:intro}Introduction}
Investigation of current fluctuations is a valuable tool for characterization of the physical mechanism of electronic transport in mesoscopic systems, in both sequential tunneling and coherent regimes. In particular, the shot noise, resulting from charge quantization, was found to provide information about the statistics and charge of quasiparticles, or the presence of interactions~\cite{blanter2000}. Due to this fact, noise has already been widely studied, both experimentally and theoretically, in the variety of systems, including multilevel quantum dots~\cite{gustavsson2006, ubbelohde2013, hasler2015, belzig2005}, quantum dot spin valves~\cite{bulka2000, weymann2008, sothmann2010}, magnetic tunnel junctions~\cite{cascales2012, cascales2014, szczepanski2016}, systems of capacitively coupled charge impurities~\cite{safonov2003}, quantum dots~\cite{schaller2010} or metallic islands~\cite{singh2016}, quantum dot molecules~\cite{kiesslich2007, sanchez2008, poltl2009}, nanoelectromechanical systems~\cite{flindt2005}, quantum dots strongly coupled to electromagnetic cavities~\cite{bergenfeldt2012}, molecules with strong electron-vibrational coupling~\cite{lau2016, koch2005}, quantum dots in the cotunneling~\cite{onac2006, okazaki2013, kaasbjerg2015} or Kondo~\cite{zarchin2008, yamauchi2011, delattre2009, ferrier2016} regimes, or systems realizing Majorana fermions~\cite{cao2012}.

Current fluctuations are most often characterized by the full counting statistics (FCS)~\cite{levitov1993, bagrets2003} which evaluates the probability distribution function of the number of charges transferred in the given time interval $\Delta t$. The studies have mainly focused on the zero-frequency FCS, which assumes $\Delta t \rightarrow \infty$ and thus analyzes the long-time behavior of current fluctuations. Less common approach, the waiting time distribution (WTD)~\cite{brandes2008}, evaluates the distribution of the time intervals separating the subsequent transport events. In contrast to the zero-frequency FCS, WTD is clearly related to the short-time dynamics of the system. For example, in some case the coherent oscillations between the quantum states of the studied system can be revealed by the clearly observable oscillatory behavior of WTD~\cite{sothmann2014, brandes2008, rajabi2013}. While FCS and WTD are mostly studied theoretically, they can be also accessed experimentally using single electron counting techniques~\cite{gustavsson2009}. For example, FCS has been used to characterize the switching between energy states of a quantum dot~\cite{fricke2007}, and WTD has been applied to investigate the spin-orbit and hyperfine interactions or charge decoherence in a double quantum dot~\cite{maisi2016, fujita2016}, or state degeneracies in a single quantum dot~\cite{hofmann2016}. It is worth mentioning that both formalisms have been used also beyond the field of electronic transport, for example in the study of quantum optical~\cite{carmichael1989, matthiesen2014, kiilerich2014, kiilerich2015} or biomolecular systems~\cite{chemla2008, moffitt2010, barato2015}.

One may ask if the complementary use of WTD and FCS gives a more complete information about the dynamics of the system, or if it sufficient to use only one of these approaches to characterize the system. In some cases, the latter conclusion may be true. For example, Bruderer~\textit{et al.}~\cite{bruderer2014} have presented a method which, in the cases considered, have enabled the full reconstruction of the generator of the dynamics of the system using the zero-frequency FCS only. As a matter of fact, when statistics of current fluctuations satisfy the renewal property, i.e. the subsequent waiting times are independent and identically distributed random variables, there exist exact identities between the cumulants of FCS and WTD~\cite{albert2011, budini2011}. As a result, $n$ cumulants of either FCS or WTD give exactly the same information. However, when subsequent waiting times are correlated the situation is different. In such a case, the cumulants of FCS and WTD become independent~\cite{albert2012} and can give complementary information about the transport mechanism. For example, my previous work~\cite{ptaszynski2017} has shown that the joint use of FCS and WTD may reduce the order of cumulants needed to fully reconstruct the generator of the dynamics of the system.

This paper goes even further -- it shows, that when the renewal property is not satisfied, it is possible that the waiting time distribution gives information about the dynamics of the system which cannot be provided by the zero-frequency full counting statistics. This conclusion is drawn from the analysis of unidirectional electron transport through a double quantum dot attached to spin-polarized leads studied by means of the quantum master equation. The analyzed system is similar to the one recently realized experimentally by F\'{a}bi\'{a}n~\textit{et al.}~\cite{fabian2016}. The studied quantum dot molecule exhibits the coherent spin oscillations in the singlet-triplet space resulting from the exchange interaction between spins. This dynamics is shown to produce the correlation between subsequent waiting times, which makes the renewal theory no longer applicable; in result WTD and the zero-frequency FCS are independent. The oscillations between the spin states result in a quite nontrivial dependence of WTD on the value of the exchange coupling and gradient of the applied magnetic field. In contrast, the zero-frequency FCS gives no insight into the internal dynamics. This fact can be intuitively explained as the result of sensitivity of the used approaches to different timescales of the dynamics: WTD is clearly related to the short-time dynamics, while the zero-frequency FCS to the long-time one.

The paper is organized as follows. Section~\ref{sec:model} specifies the analyzed model of a double quantum dot system. Section~\ref{sec:methods} presents the methods used to describe electronic transport and to calculate statistics of current fluctuations. In Sec.~\ref{sec:results} the results are presented and discussed. Finally, Sec.~\ref{sec:conclusions} brings conclusions following from my results. Appendix contains full matrix representation of the Liouvillian of the system.

\section{\label{sec:model}Model}
I consider a system of two tunnel-coupled single-level quantum dots attached to two collinearly polarized ferromagnetic leads (arranged in either parallel or antiparallel way). Similar double quantum dot molecules coupled to spin-polarized electrodes have been already examined theoretically~\cite{weymann2008, trocha2009}. In particular, the previous theoretical studies have focused on transport in the sequential tunneling~\cite{fransson2006a, fransson2006b, hornberger2008, luo2013, xue2014}, cotunneling~\cite{weymann2007} and Kondo~\cite{tanaka2004, zitko2012, krychowski2016} regimes. The analyzed system is described by the Hamiltonian which can be separated into four terms~\cite{weymann2008, hornberger2008, fransson2006a, fransson2006b}:
\begin{equation} \label{hamtotal}
\hat{H}=\hat{H}_L+\hat{H}_R+\hat{H}_{D}+ \hat{H}_T.
\end{equation}
The first two terms of the Hamiltonian describe noninteracting itinerant electrons in the left (L) and the right (R) lead. They are expressed in the following way: $\hat{H}_{\alpha}=\sum_{\mathbf{k}\sigma} \epsilon_{\alpha \mathbf{k} \sigma} c_{\alpha \mathbf{k} \sigma}^{\dagger} c_{\alpha \mathbf{k} \sigma}$, where $\alpha \in \{L,R\}$, $\epsilon_{\alpha \mathbf{k} \sigma}$ is the energy of the electron with a wave vector $\mathbf{k}$ and spin $\sigma \in \{ \uparrow,\downarrow\}$, and $c_{\alpha \mathbf{k} \sigma}^{\dagger}$ ($c_{\alpha \mathbf{k} \sigma}$) denotes the creation (annihilation) operator associated with a such electron. The third term $\hat{H}_{D}$, describing the double dot system, is given by the expression~\cite{sandalov1995, fransson2006b, fransson2006c, mireles2006}
\begin{align} \label{hamd}
\hat{H}_{D} &= \sum_{j\sigma} \epsilon d_{j\sigma}^{\dagger} d_{j \sigma}+t_{12} \sum_{\sigma} \left(d_{1\sigma}^{\dagger} d_{2\sigma}+ d_{2\sigma}^{\dagger} d_{1\sigma} \right) \\ \nonumber
&+ \sum_{j} U \hat{n}_{j \uparrow} \hat{n}_{j \downarrow}+ \left(U_{12}-\frac{J}{2} \right) \hat{n}_1 \hat{n}_2-2J \mathbf{\hat{S}}_1 \cdot \mathbf{\hat{S}_2}  \\ \nonumber
&+ \gamma (B+\Delta B) \hat{S}_1^z + \gamma (B-\Delta B) \hat{S}_2^z,
\end{align}
%
\begin{figure} 
	\centering
	\includegraphics[width=0.9\linewidth]{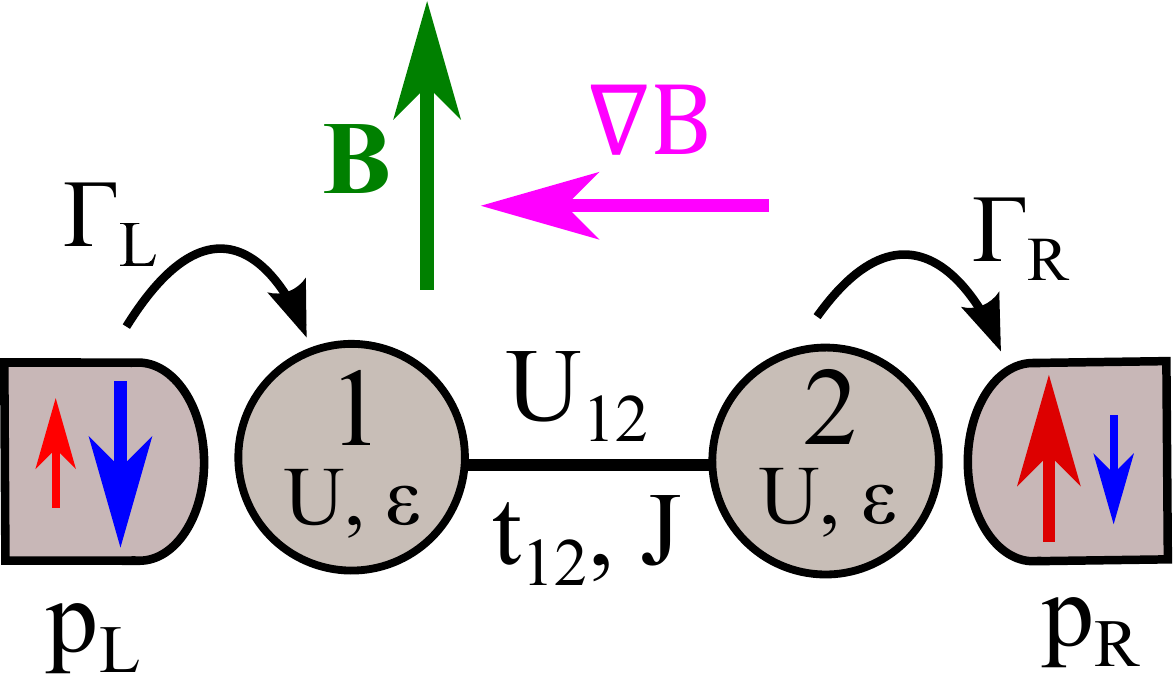} 
	\caption{Scheme of the double quantum dot attached to two spin polarized leads with spin polarization $p_L$ (left lead) and $p_R$ (right lead). Parameters $U$ and $\epsilon$ denote the intra-dot Coulomb interaction and the orbital energy respectively. $U_{12}$ -- inter-dot Coulomb interaction, $t_{12}$ -- hopping parameter, $J$ -- exchange interaction between dots, $\mathbf{B}$ -- magnetic field (with a mean value $B$), $\nabla B$ -- magnetic field gradient. Transport is unidirectional, as indicated by arrows. $\Gamma_{\alpha}$ -- mean tunneling rates (averaged over spin polarizations).}
	\label{fig:uklad}
\end{figure}
%
where $d_{j\sigma}^{\dagger}$ ($d_{j\sigma}$) is the creation (annihilation) operator of the electron with spin $\sigma$ in the first ($j=1$) or the second ($j=2$) quantum dot, $\epsilon$ is the corresponding particle energy (here assumed to be the same on both dots), and $t_{12}$ is the interdot hopping integral; $U$ and $U_{12}$ are the values of the intradot and the interdot Coulomb interaction, respectively, and particle number operators are defined as $\hat{n}_{j \sigma}=d_{j\sigma}^{\dagger} d_{j \sigma}$, $\hat{n}_{j}=\hat{n}_{j \uparrow}+\hat{n}_{j \downarrow}$. $J$ is the exchange interaction between spins in the left and the right dot, with $J>0$ ($J<0$) corresponding to the ferromagnetic (antiferromagnetic) interaction. $\mathbf{\hat{S}}_j=(1/2) \sum_{\sigma \sigma'} d_{j\sigma}^{\dagger} \boldsymbol{\sigma}_{\sigma \sigma'} d_{j\sigma'}$ denotes the spin operator, where $\boldsymbol{\sigma}=(\sigma^x,\sigma^y,\sigma^z)$ is the vector of Pauli matrices. Finally, the last two terms of $\hat{H}_{D}$ describe the Zeeman splitting, where $\gamma$ denotes the gyromagnetic ratio and $\hat{S}_j^z=(n_{j \uparrow}-n_{j \downarrow})/2$ is the operator of the spin projection on the magnetic field direction (which defines the z-axis). Here, for the sake of simplicity, the magnetic field is defined as a sum of all possible contributions, including the external magnetic field, the stray magnetic field generated by the ferromagnetic electrodes~\cite{fabian2016} or the effective magnetic field associated with the exchange coupling of the quantum dots to the leads~\cite{braun2004} (one should be aware, that the latter contribution is dependent on the parameters of the system, e.g. level positions, tunnel couplings or spin polarizations of leads). The magnetic field can be in general inhomogeneous (being equal to $B+\Delta B$ at the first and $B-\Delta B$ at the second dot), which can make the Zeeman splitting in the dots unequal. Such inhomogeneous fields with high magnetic field gradients can be created using the micromagnets~\cite{pioro2007, pioro2008, wu2014}. Alternatively, the difference of the Zeeman splittings can be generated by making the gyromagnetic ratios is the dots unequal~\cite{csonka2008, schroer2011, laucht2015}. The whole Hamiltonian $\hat{H}_D$ is quite complex, including many different parameters. However, in the later discussion the transport model will be analyzed, in which only the influence of the last three terms of the Hamiltonian $\hat{H}_D$ on transport will be important.

The term $\hat{H}_T$ describes the tunneling between the double quantum dot and the leads. It is expressed in the following way~\cite{weymann2008, hornberger2008, fransson2006a, fransson2006b}:
\begin{align}
\hat{H}_T &= \sum_{\mathbf{k}\sigma} \left(t_{L1} c_{L \mathbf{k} \sigma}^{\dagger} d_{1\sigma} +t^*_{L1} d_{1\sigma}^\dagger c_{L\mathbf{k} \sigma} \right) \\ \nonumber
&+\sum_{\mathbf{k}\sigma} \left(t_{R2} c_{R \mathbf{k} \sigma}^{\dagger} d_{2\sigma} +t^*_{R2} d_{2\sigma}^\dagger c_{R\mathbf{k} \sigma} \right),
\end{align}
where $t_{\alpha j}$ denotes the matrix element corresponding to the tunneling between the lead $\alpha$ and the $j$-th dot. The coupling strength between the $j$-th dot and the lead $\alpha$ for a specific spin polarization can be written as $\Gamma^{\sigma}_{\alpha j}=2 \pi |t_{\alpha j}|^2 \rho_{\alpha}^\sigma$, where $\rho_{\alpha}^\sigma$ denotes the spin-dependent density of states in the lead $\alpha$ (with spin polarization assumed to be parallel to the z-axis). The spin polarization of the lead $\alpha$ is defined as $p_{\alpha} = (\rho_{\alpha}^{\uparrow}-\rho_{\alpha}^{\downarrow})/(\rho_{\alpha}^{\uparrow}+\rho_{\alpha}^{\downarrow})$. Coupling strengths can be then written as $\Gamma^{\uparrow}_{\alpha j}=(1+p_\alpha) \Gamma_{\alpha j}$ and $\Gamma^{\downarrow}_{\alpha j}=(1-p_\alpha) \Gamma_{\alpha j}$, where $\Gamma_{\alpha j}=(\Gamma^{\uparrow}_{\alpha j}+\Gamma^{\downarrow}_{\alpha j})/2$. In the following part of the paper, coupling strengths are denoted in the simplified manner: $\Gamma_{L1}=\Gamma_L$, $\Gamma_{R2}=\Gamma_R$, $\Gamma_{L1}^{\sigma}=\Gamma_L^{\sigma}$, and $\Gamma_{R2}^{\sigma}=\Gamma_{R}^{\sigma}$.

One should be aware that the considered model does not include many physical effects which can be profound in real systems. For example, the model neglects the decoherence~\cite{bluhm2010} and the spin relaxation~\cite{prance2012} associated with the external electromagnetic environment, or the dependence of the tunneling rates on the spin state associated with the spatial distribution of the electron wave function~\cite{hanson2005}. However, the simplicity of the considered model facilitates the interpretation of results, which makes it a good starting point for the analysis of more realistic situations.

\section{\label{sec:methods}Methods}
Now, the methods used to describe transport through the double quantum dot molecule are presented. I focus on the transport regime in which only the single and the double occupancy of the whole system is allowed. To achieve such a regime, the parameters are assumed to fulfill the following conditions: ${\epsilon = -U_{12}}$, $U_{12} \gg k_B T$, $\mu_L-2|t_{12}|-\gamma {(|B|+|\Delta B|)}-|J| \gg k_B T$, $U-\mu_L-2|t_{12}|-\gamma {(|B|+|\Delta B|)}-|J| \gg k_B T$, $-\mu_R-2|t_{12}|-\gamma {(|B|+|\Delta B|)}-|J| \gg k_B T$ and $U+\mu_R-2|t_{12}|-\gamma {(|B|+|\Delta B|)}-|J| \gg k_B T$ where $\mu_L$ ($\mu_R$) is the electrochemical potential of the left (right) lead. Moreover, it is assumed that the single dot can be occupied by at most one electron. Although for a finite value of the exchange coupling $J$ there is always a finite probability of a double occupancy of a single dot~\cite{hu2000, schliemann2001}, for sufficiently low values of $J$ it can be safely neglected in the discussion of transport. Thus, there are eight quantum states within the transport window, which in the localized basis can be denoted as $\{{|\! \! \uparrow \! \!  0 \rangle}, {|\! \!  \downarrow \! \! 0 \rangle}, {|0 \! \! \uparrow \rangle}, {|0 \! \! \downarrow \rangle}, {|\! \! \uparrow \uparrow \rangle}, {|\! \! \uparrow \downarrow \rangle}, {|\! \! \downarrow \uparrow \rangle}, {|\! \! \downarrow \downarrow \rangle\}}$, where the first and the second position in the ket correspond to the first and the second dot, respectively; arrows denote the spin orientation of the occupying electron and 0 denotes the absence of the electron. Furthermore, since the separation of the electrochemical potentials of the leads from the relevant energy levels of the double quantum dot system is assumed to be large in comparison with $k_B T$, thermally excited tunneling against the bias can be neglected and transport can be considered as unidirectional. In such a case, in the regime of the weak coupling between the dots and the leads, transport can be well described by the Markovian master equation~\cite{gurvitz1996, gurvitz1998}. In the localized basis one obtains the generalized master equation (GME)~\cite{poltl2009} which describes the time dependence of the density matrix of the system. It can be written in the Lindblad form
\begin{align} \label{lindblad}
& \frac{d \rho}{dt} = -i \left[ \hat{H}_D,\rho \right] \\ \nonumber
&+ \sum_{\sigma, \sigma'} \frac{\Gamma_L^{\sigma}}{2} \left(2 l^{\dagger}_{\sigma \sigma'} \rho l_{\sigma \sigma'}-l_{\sigma \sigma'} l^{\dagger}_{\sigma \sigma'} \rho -\rho l_{\sigma \sigma'} l^{\dagger}_{\sigma \sigma'} \right) \\ \nonumber
&+ \sum_{\sigma, \sigma'} \frac{\Gamma_R^{\sigma'}}{2} \left(2 r^{\dagger}_{\sigma \sigma'} \rho r_{\sigma \sigma'}- r_{\sigma \sigma'} r^{\dagger}_{\sigma \sigma'} \rho -\rho r_{\sigma \sigma'} r^{\dagger}_{\sigma \sigma'} \right),
\end{align}
where $\rho$ is the density matrix and the Lindblad operators describing the tunneling through the left and right junction, respectively, are defined as $l^{\dagger}_{\sigma \sigma'}=|\sigma \sigma' \rangle \langle 0 \sigma'|$ and $r^{\dagger}_{\sigma \sigma'}=|\sigma 0 \rangle \langle \sigma \sigma'|$. Here, and in the whole paper, I take $\hbar=1$.

Finally, I present the methods used to describe the waiting time distribution and the full counting statistics. The description largely follows the one included in my previous paper \cite{ptaszynski2017}, and the reader is referred to it for a more detailed discussion. The master equation is now written in the Liouville space~\cite{carmichael1993, breuer2002}:
\begin{equation} \label{mastereq}
\dot{\rho}(t)=\mathcal{L} \rho(t),
\end{equation}
where $\rho(t)$ is the state vector in the Liouville space, which contains both diagonal and off-diagonal elements of the density matrix of the system (i.e. the state probabilities and the coherences), and $\mathcal{L}$ is the square matrix representing the Liouvillian. Full matrix forms of $\rho(t)$ and $\mathcal{L}$ are presented in Eqs.~\eqref{fullvec}-\eqref{fullliov} in the Appendix.

To calculate the waiting time distribution the approach of Brandes~\cite{brandes2008} is used. The distribution of waiting times between jumps of types $k$ and $l$ is denoted as $w_{kl}(\tau)$. Here the term ``jump of type $k$" refers to the specific tunneling process: the subscript $L$ is related to the electron jump from the left lead to the first dot, and the subscript $R$ is associated with the tunneling from the second dot to the right lead. It is assumed, that the counting procedure do not distinguish the spin of the transferred electrons, which corresponds to standard single charge counting experiments. One can consider distributions of waiting times either between subsequent tunnelings through the same junction [$w_{LL}(\tau)$ and $w_{RR}(\tau)$], or between jumps through different junctions [$w_{LR}(\tau)$ and $w_{RL}(\tau)$]. The Laplace transform of the distribution $w_{kl}(\tau)$ reads~\cite{brandes2008}
\begin{equation} \label{wkldis}
w_{kl}(s)=\int_0^{\infty} e^{-s \tau} w_{kl}(\tau) d\tau=\frac{\text{Tr}[\mathcal{J}_l (s-\mathcal{L}_0^{kl})^{-1} \mathcal{J}_k \rho_0]}{\text{Tr}[\mathcal{J}_k \rho_0]},
\end{equation}
where $k,l \in \{L,R \}$. $\mathcal{J}_k$ and $\mathcal{J}_l$ are the operators containing all off-diagonal elements of the Liouvillian corresponding to the jumps through the $k$ and the $l$ junction, respectively. Elements of the Liouvillian included in the jump operators are explicitly presented in Eq.~\eqref{fullliov} in the Appendix. The more elaborate discussion of the construction of the jump operators can be found in the paper of Brandes~\cite{brandes2008}. The operator $\mathcal{L}_0^{kl}=\mathcal{L}-\mathcal{J}_k-(1-\delta_{kl}) \mathcal{J}_l$ is the reminder of the Liouvillian, and $\rho_0$ is the vector of the stationary state (the solution of the equation $\mathcal{L} \rho=0$). The distribution $w_{kl}(\tau)$ is the inverse Laplace transform of $w_{kl}(s)$. It is often convenient to consider cumulants of the distribution rather than the whole distribution itself. The cumulants of WTD can be calculated using the formula~\cite{brandes2008}
\begin{align} \label{cumst}
\kappa_n^{kl} = (-1)^n \left \{\frac{d^n \ln[w_{kl}(s)]}{ds^n} \right \}_{s=0},
\end{align}
where  $\kappa_n^{kl}$ is the $n$th order cumulant of the distribution $w_{kl}(\tau)$.

Additionally, the useful information about the transport mechanism can be provided by the joint distribution of two subsequent waiting times $\tau_{kl}$ and $\tau_{lm}$ (with $k,l,m \in \{L,R \}$), denoted as $w_{klm} (\tau_{kl}, \tau_{lm})$~\cite{ptaszynski2017}. Its Laplace transform is defined as~\cite{ptaszynski2017}
\begin{align}
& w_{klm}(s,z) \\ &= \nonumber \int_0^\infty d\tau_{lm} \int_0^\infty d\tau_{kl} e^{-s \tau_{kl}-z \tau_{lm}} w_{klm}(\tau_{kl},\tau_{lm}) \\ \nonumber &= 
\frac{\text{Tr}[\mathcal{J}_m (z-\mathcal{L}_0^{lm})^{-1} \mathcal{J}_l (s-\mathcal{L}_0^{kl})^{-1} \mathcal{J}_k \rho_0]}{\text{Tr}[\mathcal{J}_k \rho_0]}.
\end{align}
This distribution can be used to calculate the cross-correlation (covariance) of two successive waiting times~\cite{ptaszynski2017}
\begin{align} \label{crosscor}
\langle \Delta \tau_{kl} \Delta \tau_{lm} \rangle = \left \{ \frac{\partial }{\partial s} \frac{\partial }{\partial z} \ln[w_{klm} (s,z)]\right\}_{s=0,z=0},
\end{align}
where $\Delta \tau_{ij}=\tau_{ij}- \langle \tau_{ij} \rangle$ is the deviation from the mean. It is convenient to analyze the normalized cross-correlation~\cite{ptaszynski2017} (i.e. the Pearson correlation coefficient~\cite{sirca2016})
\begin{align} \label{ncc}
\text{NCC}=\frac{\langle \Delta \tau_{kl} \Delta \tau_{lm} \rangle}{\sqrt{\langle \Delta \tau_{kl}^2 \rangle \langle \Delta \tau_{lm}^2 \rangle}},
\end{align}
where $\langle \Delta \tau_{ij}^2 \rangle = \langle \tau_{ij}^2 \rangle-\langle \tau_{ij} \rangle^2=\kappa_2^{ij}$ is the variance of the waiting time distribution. It has a value between $-1$ and $1$, where the former (latter) limit is associated with the perfect negative (positive) correlation~\cite{sirca2016}. The equality ${\text{NCC}=0}$ is the necessary (however not sufficient) condition of the statistical independence of the subsequent waiting times~\cite{sirca2016, ptaszynski2017}.

Finally, the full counting statistics is considered. FCS analyzes the distribution of the number of particles tunneling through the single junction in a given time interval $\Delta t$. I focus on the zero-frequency FCS which assumes $\Delta t \rightarrow \infty$. In such a case, distributions measured at the left and the right junction are the same due to Kirchoff's current law. The calculated quantities are the $n$th order scaled cumulants of the FCS, defined in the following way: $c_n = \lim_{\Delta t \to \infty} C_n(\Delta t)/\Delta t$, where $C_n(\Delta t)$ is the $n$th order cumulant of the number of jumps through a single junction taking place in the time interval $\Delta t$. Scaled cumulants up to the $N$th order can be calculated using the following set of equations~\cite{bruderer2014, wachtel2015}:
\begin{align} \label{kumkal}
&\left\{ \frac{d^n}{d \chi^n} \det \left[ \lambda(\chi)-\mathcal{L}_0-\mathcal{J} e^{\chi} \right] \right\}_{\chi = 0}= 0,  \\ \nonumber
& c_n=\left[ \frac{d^n}{d\chi^n} \lambda(\chi)\right]_{\chi = 0},
\end{align}
where $n$ ranges from $1$ to $N$, $\chi$ is a counting field, $\lambda(\chi)$ is a scaled cumulant generating function with the property $\lambda(0)=0$, $\mathcal{J}$ is either $\mathcal{J}_L$ or $\mathcal{J}_R$, and $\mathcal{L}_0=\mathcal{L}-\mathcal{J}$. Since $\lambda(\chi=0)=0$, knowledge of the exact form of $\lambda(\chi)$ is not necessary for the calculation. A more elaborate discussion of FCS formalism can be found in Refs.~\cite{bagrets2003, esposito2009, touchette2009}.


\section{\label{sec:results}Results}
In this section I analyze the influence of the coherent oscillations between the spin states of the molecule on current fluctuations. Such oscillations take place because the electron tunneling may initialize the states ${|\! \! \uparrow \downarrow \rangle}$ and ${|\! \! \downarrow \uparrow \rangle}$ which, in general, are not eigenstates of the Hamiltonian $\hat{H}_D$ but their superpositions. The eigenstates corresponding to the zero spin z-component read as
 %
 \begin{figure} 
 	\centering
 	\includegraphics[width=0.8\linewidth]{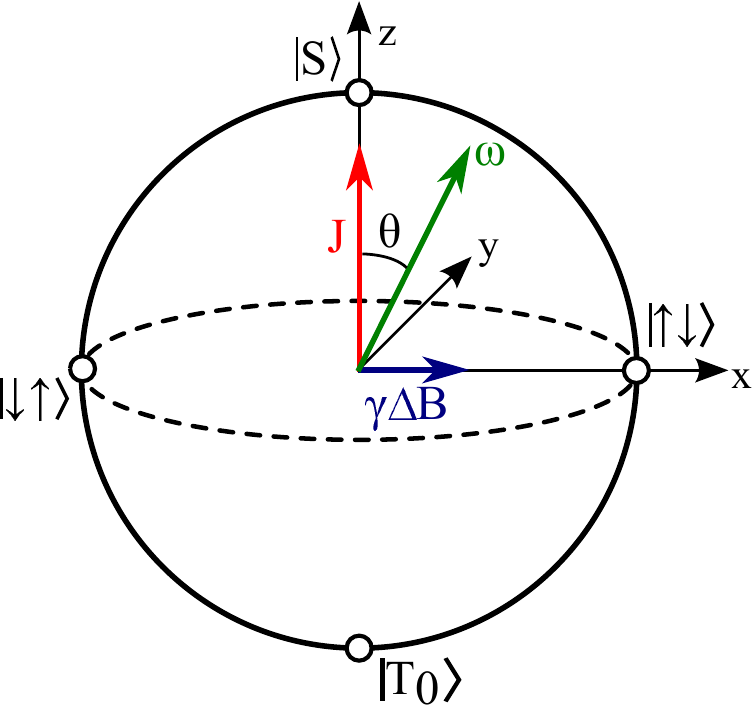} 
 	\caption{Scheme of the Bloch sphere of the qubit in the $S-T_0$ space. The z-axis is defined as parallel to the vectors describing the states $|S \rangle$ and $|T_0 \rangle$, and the x-axis is parallel to the vectors of the ${|\! \! \uparrow \downarrow \rangle}$ and ${|\! \! \downarrow \uparrow \rangle}$ states. The rotation axis denotes as $\boldsymbol{\omega}$ is located in the xz plane, forming an angle $\theta$ with the z-axis.}
 	\label{fig:bloch}
 \end{figure}
 %
\begin{align} \label{psi1}
{| \psi_P \rangle} &= \frac{1}{\sqrt{2}} \left[ \cos \left(\frac{\theta}{2} \right)  {|S \rangle}+ \sin \left(\frac{\theta}{2} \right)  {|T_0 \rangle} \right], \\ \label{psi2}
{| \psi_A \rangle} &= \frac{1}{\sqrt{2}} \left[- \sin \left(\frac{\theta}{2} \right)  {|S \rangle}+\cos \left(\frac{\theta}{2} \right)  {|T_0 \rangle} \right].
\end{align}
where $|S \rangle = {(|\! \! \uparrow \downarrow \rangle-|\! \! \downarrow \uparrow \rangle)/\sqrt{2}}$ and $|T_0 \rangle={(|\! \! \uparrow \downarrow \rangle+|\! \! \downarrow \uparrow \rangle)/\sqrt{2}}$ are the singlet and the triplet states, respectively, and $\theta$ is an angle defined by following equations: $\cos \theta = J/\sqrt{J^2+(\gamma \Delta B)^2}$ and $\sin \theta = \gamma \Delta B/\sqrt{J^2+(\gamma \Delta B)^2}$. For $\theta=0$ these eigenstates are equivalent to $|S \rangle$ and $|T_0 \rangle$ states, respectively, and for $\theta=\pi/2$ they are equivalent to the ${|\! \! \uparrow \downarrow \rangle}$ and ${|\! \! \downarrow \uparrow \rangle}$ states. Dynamics of the spin states can be described as a qubit rotation on the Bloch sphere (Fig.~\ref{fig:bloch}), with the direction of the rotation axis defined by the angle $\theta$ and the angular frequency of the rotation being equal to $\omega=\sqrt{J^2+(\gamma \Delta B)^2}$. The states $| \psi_P \rangle$ and ${| \psi_A \rangle}$ are then associated with vectors parallel and antiparallel to the qubit rotation axis. For $\theta=0$ the qubit rotates along the equator of the Bloch sphere which corresponds to the coherent oscillations between the states ${|\! \! \uparrow \downarrow \rangle}$ and ${|\! \! \downarrow \uparrow \rangle}$. One may note a resemblance of the considered dynamics to the principle of operation of the singlet-triplet qubits~\cite{petta2005, maune2011, kawakami2014, wu2014, bluhm2010, prance2012}. It should be mentioned, that the influence of the coherent oscillations between the spin states on transport through the double quantum dot system has been already studied by S\'{a}nchez \textit{et al.}~\cite{sanchez2008b}. However, in this paper both the mechanism generating the oscillations and the transport regime (associated with the choice of the system parameters) are significantly different from the ones considered therein.

The section is divided into two subsections, analyzing the cases of the homogeneous (Sec.~\ref{subsec:gb0}) and the inhomogeneous magnetic field (Sec.~\ref{subsec:gbpos}), respectively. For the homogeneous magnetic field it will be shown that the waiting time distribution exhibits the nontrivial dependence on the value of the exchange coupling between dots while the zero-frequency full counting statistics is independent of this parameter. Since this dependence is observable only for sufficiently low values of $J$, in the next subsection I consider the case of the inhomogeneous magnetic field in which the dependence of WTD on the direction of the rotation axis can be observed also for high values of $J$.

\subsection{\label{subsec:gb0}$\nabla B=0$}
 
I begin my analysis by considering the case when the external magnetic field is homogeneous ($\nabla B=0$), and therefore the states ${|\! \! \uparrow \downarrow \rangle}$ and ${|\! \! \downarrow \uparrow \rangle}$ are degenerate. To simplify the analysis I assume that $|t_{12}| \gg |J|, \Gamma_L, \Gamma_R$. In such a case, due to large value of the inter-dot tunnel coupling, the coherent oscillations between the states $|\sigma 0\rangle$ and $|0 \sigma \rangle$ are so fast, that these states are non-distinguishable in timescales corresponding to the tunneling between the leads and the dots. Therefore, for the sake of simplicity of the qualitative discussion of the transport mechanism, one may effectively reduce the rank of the density matrix of the system by replacing four diagonal elements corresponding to the dot-localized states ($|\sigma 0\rangle$ and $|0 \sigma \rangle$) with two elements associated with the subspaces of states corresponding to different spin orientations: ${|\! \uparrow \rangle} \equiv \{{|\! \uparrow 0\rangle}, {|0 \! \uparrow \rangle} \}$ and ${|\! \downarrow \rangle} \equiv \{ {|\! \downarrow 0\rangle}, {|0 \! \downarrow \rangle} \}$. It can be done because the probabilities of these states and subspaces are strictly related [$P(|\sigma \rangle)=2P(|0\sigma \rangle)=2P(|\sigma 0 \rangle )$], and therefore the master equation can be easily rewritten to describe the dynamics in the reduced basis. The reduced Liouvillian of the system is presented in Eq.~\eqref{liovred} in the Appendix. Figure~\ref{fig:6statescoh}~(a) shows the graphical representation of the model of the dynamics of the system obtained after the reduction of the basis.

%
\begin{figure}
	\centering
	\subfloat[]{\includegraphics[width=0.7\linewidth]{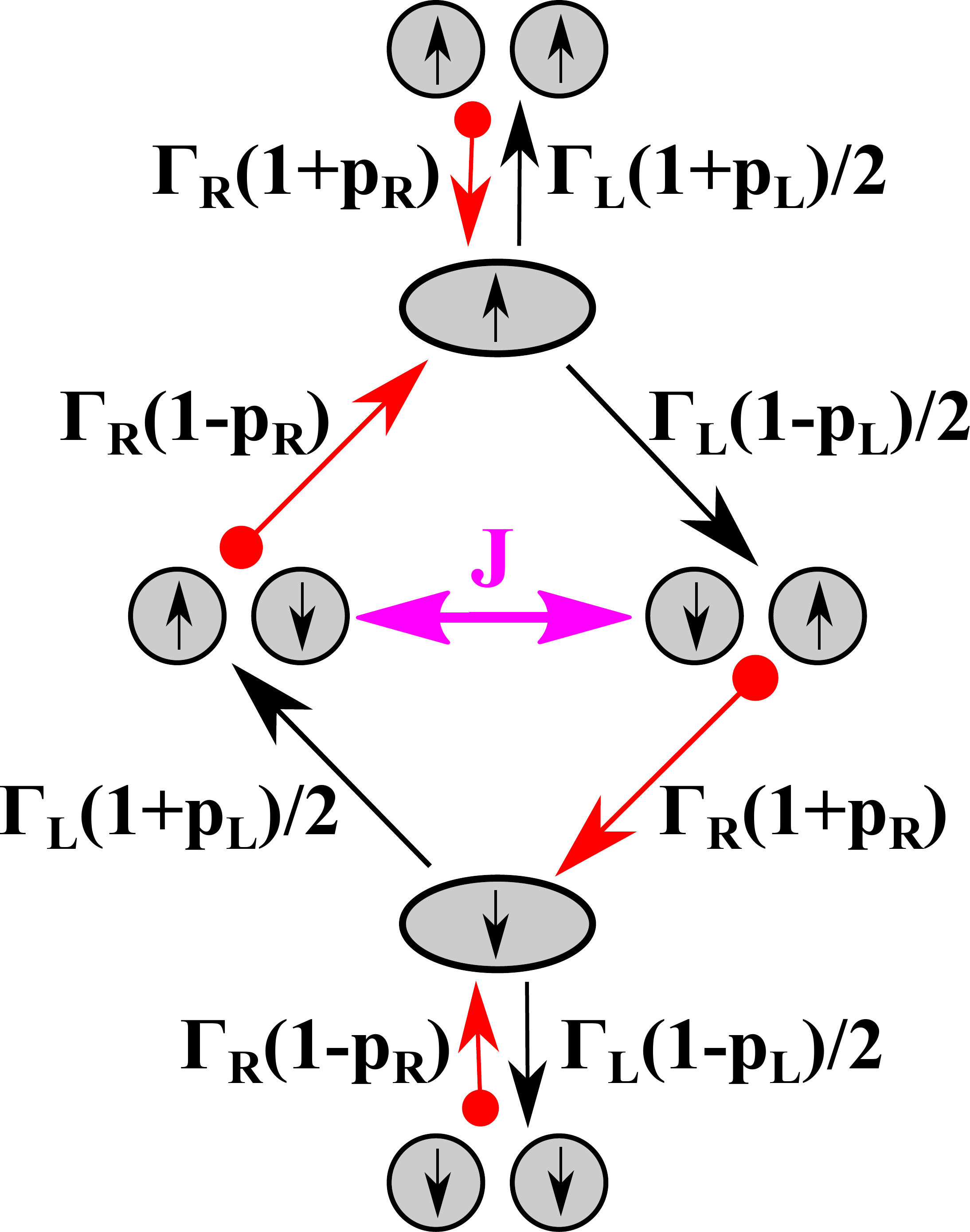}} \\
	\subfloat[]{\includegraphics[width=0.9455\linewidth]{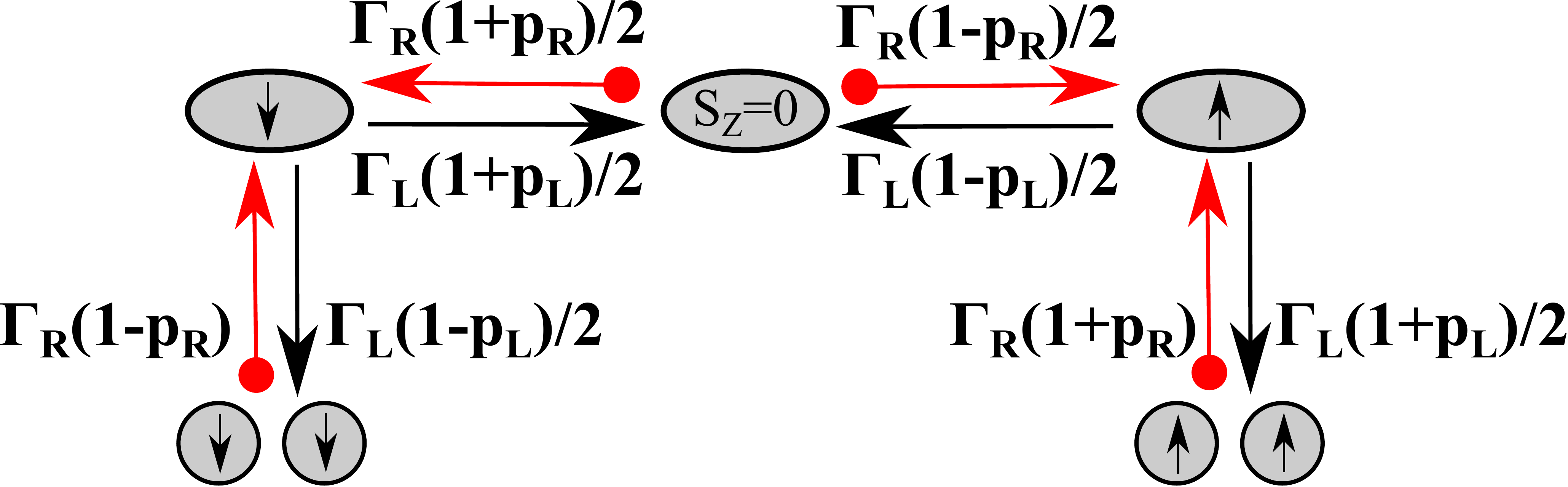}}
	\caption{(a) Six-state model of the dynamics of the system for $\nabla B = 0$. Red arrows with bullet tails correspond to the tunneling from the molecule to the right lead, while black arrows to the tunneling from the left lead to the molecule. The violet double-ended arrow in the middle denotes coherent oscillations between states. The tunneling rates $\Gamma_L^\sigma$ are divided by 2 since the electron can tunnel to the molecule only when the left dot is empty (state $|0 \sigma \rangle$ is occupied). (b) Simplified five-state model for $J \gg \Gamma_L, \Gamma_R$. ${|S_z=0\rangle}$ is the subspace containing the states  ${|\! \! \uparrow \downarrow \rangle}$ and ${|\! \! \downarrow \uparrow \rangle}$.}
	\label{fig:6statescoh}
\end{figure}
%

At first, WTD is analyzed. Let us start with the distribution of waiting times between the jump from the molecule to the right lead and the successive jump from the left lead to the molecule, denoted as $\tau_{RL}$. This distribution, describing the filling of the singly-occupied molecule by the second electron, has a simple exponential form: $w_{RL}(\tau)=\Gamma_{L} e^{-\Gamma_{L} \tau}$. The mean time $\langle \tau_{RL} \rangle$ is equal to $1/\Gamma_L$. The distribution depends only on the parameter $\Gamma_L$ because the rate of electron tunneling to the left dot does not depend on the spin state of the electron occupying the molecule. It is independent of the polarization of the left lead, which influences only the proportion of the number of $\uparrow$ and $\downarrow$ spins tunneling to the molecule.
%
\begin{figure} 
	\centering
	\includegraphics[width=0.9\linewidth]{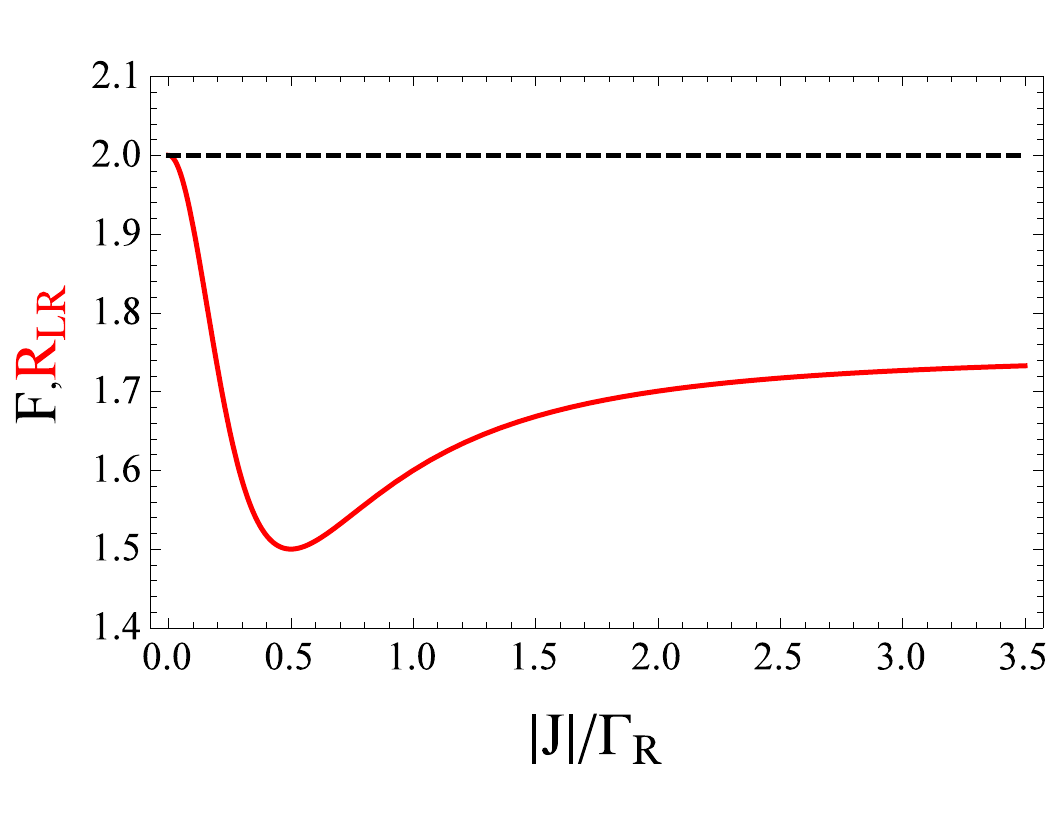} 
	\caption{Dependence of the Fano factor $F$ (black dashed line) and the randomness parameter $R_{LR}$ (red solid line) on $|J|/\Gamma_R$ for $p_L=0$ and $p_R=\sqrt{2}/2 \approx 0.7$. Fano factor calculated in the limit of $\Gamma_L \gg \Gamma_R$.}
	\label{fig:fanorandkoh}
\end{figure}
%

The distribution of the waiting times between the jump from the left lead to the first dot and the successive jump from the second dot to the right lead, denoted as $\tau_{LR}$, is much more complex since the rate of electron tunneling to the right lead depends on the spin state of the molecule. The formula describing the waiting time distribution is in general complicated and depends on the parameters $\Gamma_R$, $p_L$, $p_R$ and $J$. In particular, dependence on $J$ shows the influence of the coherent oscillations between the spin states. The mean waiting time $\langle \tau_{LR} \rangle$, however, is independent of $J$ and equal to $(1- p_L p_R)/\Gamma_R (1-p_R^2)$. This may be nonintuitive, since the coherent oscillations cause the spin-flip in the right dot, which should affect the rate of electron tunneling. However, it appears that two opposite spin-flip processes ($\uparrow \rightarrow \downarrow$ and $\downarrow \rightarrow \uparrow$) equilibrate themselves, thus leaving the spin population of the right dot intact. 

The dependence of the shape of WTD of $J$ can be, however, observed when one analyzes the higher cumulants of the distribution. In this work I will focus on the randomness parameter $R_{LR}$, associated with the second cumulant, which characterizes the spread of the distribution around the mean value. It is defined as follows
\begin{align} \label{rand}
R_{LR}=\frac{\kappa^{LR}_2}{(\kappa^{LR}_1)^2}=\frac{\langle \Delta \tau_{LR}^2 \rangle}{\langle \tau_{LR} \rangle^2},
\end{align}
where $\kappa^{LR}_n$ are the cumulants of $w_{LR}(\tau)$ [see Eq.~\eqref{cumst}] and $\langle \Delta \tau_{LR}^2 \rangle$ is the variance of this distribution. This parameter, similarly to the distribution $w_{LR}(\tau)$, is dependent on $J$, as well as on parameters $\Gamma_R$, $p_L$, and $p_R$. For $p_L=0$ it is given by the expression
\begin{align} \label{rlrpl0}
R_{LR}=1 + 2 p_R^2 \left[1  \right. &+ \frac{4J^2}{2 J^2 + (1 - p_R^2) \Gamma_R^2}  \\ \nonumber &- \left. \frac{2 J^2 (5 - p_R^2)}{4 J^2 + (1 - p_R^2) \Gamma_R^2} \right].
\end{align}
The red solid line in Fig.~\ref{fig:fanorandkoh} presents the dependence of the randomness parameter on the ratio $|J|/\Gamma_R$ (which can be tuned by the gate voltages~\cite{fabian2016, fujiwara2006}) for $p_L=0$ and $p_R=\sqrt{2}/2 \approx 0.7$ (it is independent of the sign of $J$). The qualitative character of the dependence of $R_{LR}$ on $|J|/\Gamma_R$ appears to be independent of the spin polarizations of the leads (assuming $p_R \neq 0$). For $J=0$ the randomness parameter is maximized. With the increase of $|J|$, $R_{LR}$ decreases, reaching the minimum for a certain finite value of $|J|/\Gamma_R$. As $|J|$ rises further, $R_{LR}$ tends asymptotically to some finite value which is lower than the one corresponding to $J=0$.
%
\begin{figure}
	\centering
	\subfloat{\includegraphics[width=0.9455\linewidth]{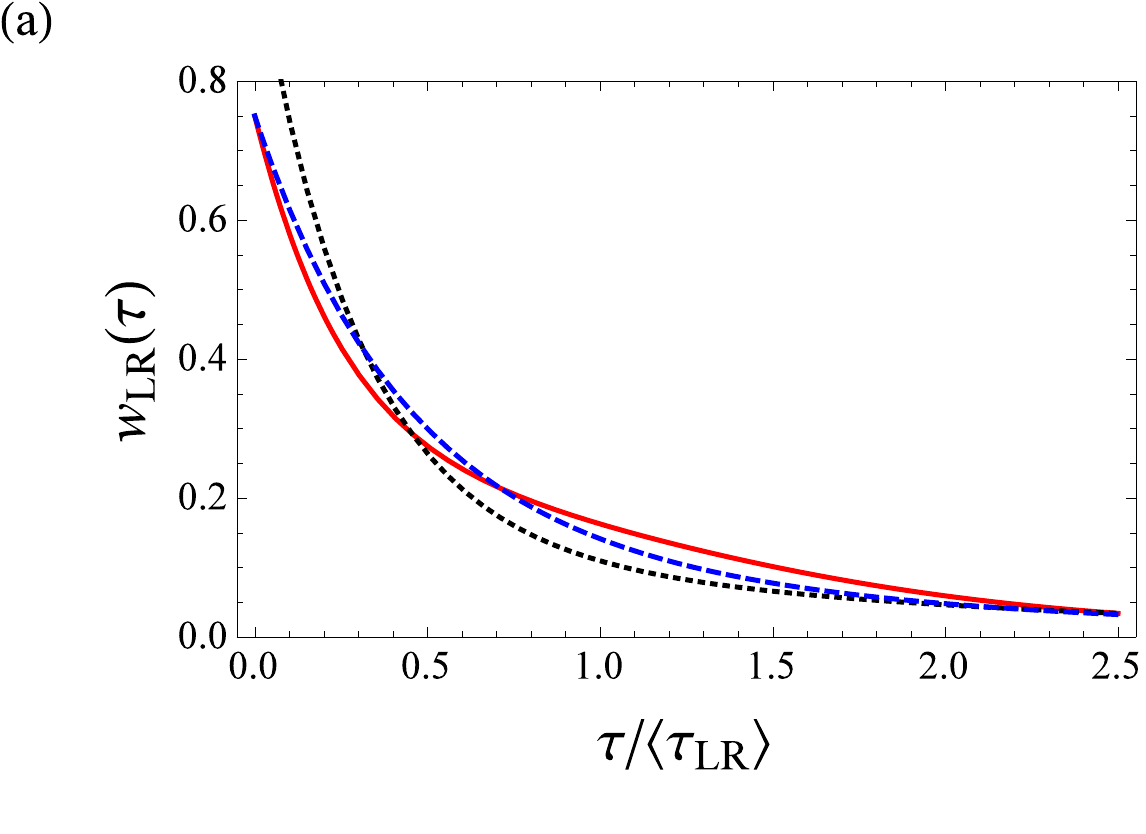}} \\
	\subfloat{\includegraphics[width=0.9455\linewidth]{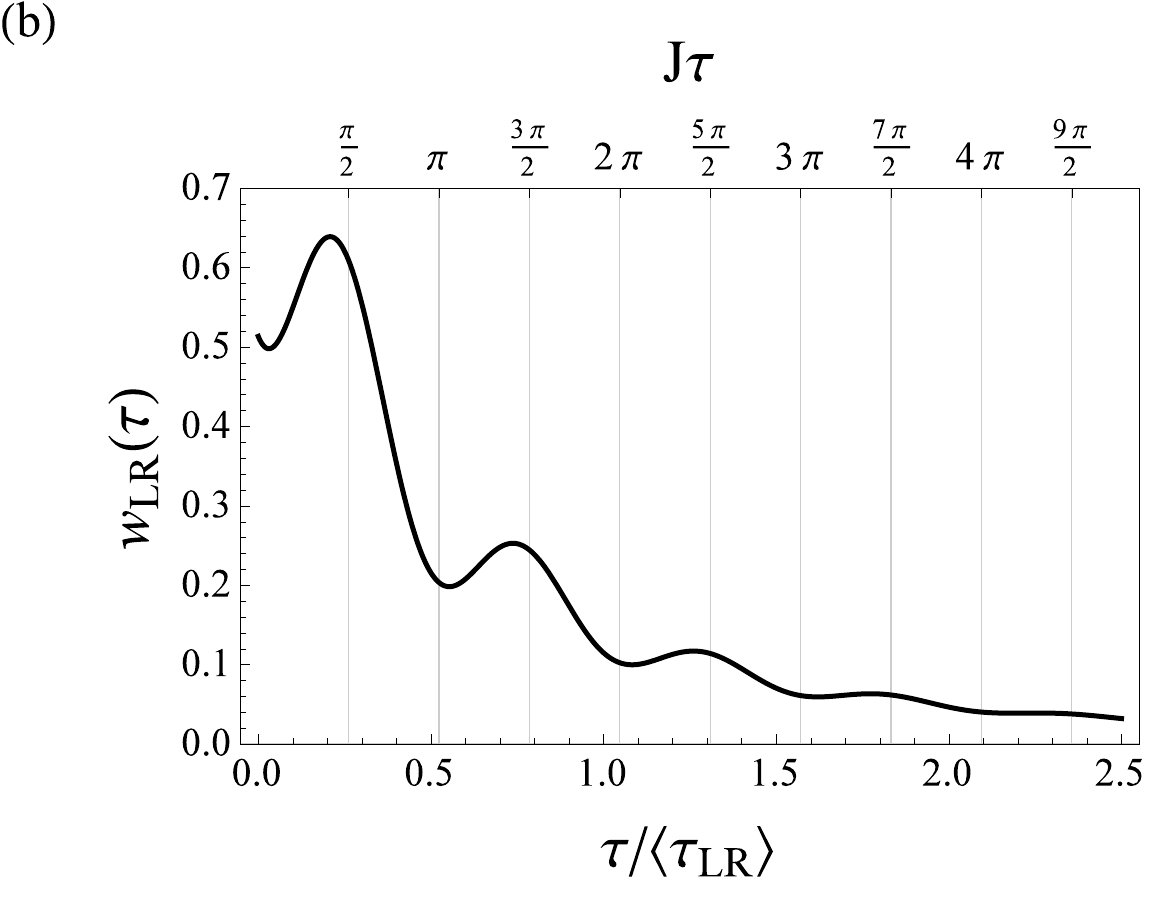}}
	\caption{(a) Waiting time distribution $w_{LR}(\tau)$ for $|J|=0$ (black dotted line), $|J|/\Gamma_R=1/2$ for which $R_{LR}$ is minimized (red solid line) and $|J| \gg \Gamma_R$ (blue dashed line). (b) Oscillatory behavior of $w_{LR}(\tau)$ for $|J|/\Gamma_R=3$. All results for $p_L=0$ and $p_R=\sqrt{2}/2 \approx 0.7$.}
	\label{fig:wtds}
\end{figure}
%

To explain this behavior, let us focus on three characteristics points: $J=0$, $|J| \gg \Gamma_R$ and the minimum of $R_{LR}$ [see Fig.~\ref{fig:wtds}~(a)]. Since the polarization of the left lead influences the results only quantitatively and does not lead to qualitatively new effects, I assume that $p_L=0$. The waiting time distribution $w_{LR}(\tau)$ for $J=0$ is given by the double-exponential function
\begin{align} \label{wj0}
& w_{LR}(\tau,p_L=0,J=0) \\ \nonumber & = \frac{1}{2} \Gamma_R \left[ (1+p_R) e^{-(1+p_R)\Gamma_R \tau}+ (1-p_R) e^{-(1-p_R)\Gamma_R \tau} \right].
\end{align}
It is a sum of two separate exponential distributions corresponding to different spin transport channels, which is associated with the fact, that electrons with spins $\uparrow$ and $\downarrow$ preserve their spins for the whole time during which they occupy the molecule (due to $J=0$). Since the timescales of the tunneling for $\uparrow$ and $\downarrow$ spins differ, the waiting times tends to deviate from the mean value. This explains the relatively high value the randomness parameter.

Next, I analyze the case of $|J| \gg \Gamma_R$. In such a case the oscillations between the states ${|\! \! \uparrow \downarrow \rangle}$ and ${|\! \! \downarrow \uparrow \rangle}$ are so fast (relatively to the timescale of the tunneling through the right lead), that distinctive character of these states is not observed. Therefore, one may further reduce the rank of the density matrix by replacing the diagonal elements corresponding to the states ${|\! \! \uparrow \downarrow \rangle}$ and ${|\! \! \downarrow \uparrow \rangle}$ by a single element corresponding to the subspace $|S_z=0 \rangle \equiv \{ {|\! \! \uparrow \downarrow \rangle}, {|\! \! \downarrow \uparrow \rangle} \}$ [with the probability ${P(|S_z=0 \rangle)}={2P(|\! \! \uparrow \downarrow \rangle)}={2P(|\! \! \downarrow \uparrow \rangle)}$]. In this way one obtains the effective five-state model of the dynamics of the system [see Fig.~\ref{fig:6statescoh}~(b)]. The waiting time distribution is now given by the triple-exponential function
\begin{align} \label{wjinf}
w_{LR}(\tau ,p_L=0,&|J| \gg \Gamma_R)   
\\ \nonumber = \Gamma_R \left[\frac{1}{2} e^{-\Gamma_R \tau} \right. &+ \frac{1-p_R}{4} (1+p_R) e^{-(1+p_R)\Gamma_R \tau} 
\\ \nonumber  &+ \left. \frac{1+p_R}{4} (1-p_R) e^{-(1-p_R)\Gamma_R \tau} \right],
\end{align}
in which the successive terms are associated with different transport channels. The first term of WTD is related to the tunneling from the $|S_z=0 \rangle$ subspace, and is expressed as the product of the probability that this subspace is generated after the tunneling from the left lead, equal to 1/2, and the exponential term describing the distribution of waiting times corresponding to the tunneling in this transport channel only. The exponential term decays with a rate equal to the mean tunneling rate $\Gamma_R$, which is associated with the mixing of states ${|\! \! \uparrow \downarrow \rangle}$ and ${|\! \! \downarrow \uparrow \rangle}$. Analogously, the second term of WTD is a product of the probability that the state ${|\! \! \uparrow \uparrow \rangle}$ is generated after the tunneling from the left lead, equal to $(1-p_R)/4$, and the exponential term $(1+p_R)\Gamma_R e^{-(1+p_R) \Gamma_R \tau}$ [here the factor $(1+p_R)\Gamma_R$ before the exponent is a normalizing constant]. As one can note, for $p_R>0$ the probability of generation of the state ${|\! \! \uparrow \uparrow \rangle}$ is decreased. It is due to the accumulation of the spin $\downarrow$ electrons in the right dot, which results from the fact that tunneling from the $|S_z=0 \rangle$ subspace produces the ${|\! \! \downarrow \rangle}$ rather than the ${|\! \! \uparrow \rangle}$ subspace [cf. Fig.~\ref{fig:6statescoh}~(b)]. The third term can be described in a way analogous to the former. As Fig.~\ref{fig:wtds}~(a) shows, the probability density of very short waiting times (close to 0) is decreased in comparison with the case analyzed previously, while for waiting times close to the mean value it is increased. This explains the smaller value of the randomness parameter. 

For intermediate values of $|J|$ the formula describing the waiting time distribution is complicated. Let us describe qualitatively the case of the minimal value of $R_{LR}$ [the red line in Fig.~\ref{fig:wtds}~(a)], assuming $p_R>0$. For $\tau=0$ the probability density $w_{LR}(0)$ is more or less similar to the one corresponding to $|J| \gg \Gamma_R$ [in a specific case shown in Fig.~\ref{fig:wtds}~(a) they are equal]. As previously, this results from the decreased population of spin $\uparrow$ electrons in the right dot. However, in contrast to the case of $|J| \gg \Gamma_R$, the frequency of oscillations is now comparable with the timescale of the tunneling through the right lead. Thus, populations of states ${|\! \! \uparrow \downarrow \rangle}$ and ${|\! \! \downarrow \uparrow \rangle}$ are not immediately equilibrated after the tunneling from the left lead, but the probability of the state  ${|\! \! \uparrow \downarrow \rangle}$ remains larger for some finite time. Due to this fact, $w_{LR}(\tau)$ decreases below the value corresponding to $|J| \gg \Gamma_R$ for small values of $\tau$. For $\tau$ comparable with a mean waiting time $\langle \tau_{LR} \rangle$ the population of the state ${|\! \! \downarrow \uparrow \rangle}$ is, on the other hand, increased due to the qubit rotation. This results in the rise of $w_{LR}(\tau)$ to values greater than the ones observed for $|J| \gg \Gamma_R$. In consequence, WTD is more concentrated around the mean value, which results in the decrease of the randomness parameter. 

In Fig.~\ref{fig:wtds}~(a) the oscillations between the spin states are not clearly visible: for $J=0$ they do not occur, for $|J| \gg \Gamma_R$ they are too fast to be observed, and for $|J|/\Gamma_R=1/2$ WTD decays with a rate faster than the frequency of the oscillations. The oscillatory behavior of the waiting time distribution, resulting from the oscillations between the spin states, can be however clearly seen for moderately high values of $|J|/\Gamma_R$, as shown in Fig.~\ref{fig:wtds}~(b). It can be interpreted as follows: For $\tau=0$ the population of the state ${|\! \! \uparrow \downarrow \rangle}$ is higher than of the state ${|\! \! \downarrow \uparrow \rangle}$, and therefore the probability of tunneling is relatively low. For $J \tau=\omega \tau \approx (n+1/2) \pi$ (where $\omega$ is the angular frequency of oscillations and $n$ is an integer) the spin states are reversed, and WTD reaches the local maxima (the maxima are slightly shifted due to the presence of the exponential decay). Conversely, for $J \tau \approx n \pi$ the state ${|\! \! \uparrow \downarrow \rangle}$ is again more populated and WTD reaches the local minima. 

Let us now analyze the zero-frequency full counting statistics, which describes the long time behavior of current fluctuations. I focus on the second-order cumulant by analyzing the Fano factor
\begin{align}  
F=\frac{c_2}{c_1}=\lim_{t \to \infty} \frac{\langle \left[\Delta n(\Delta t) \right]^2 \rangle}{\langle n(\Delta t) \rangle}, 
\end{align}
where $c_n$ are the cumulants of FCS [see Eq.~\eqref{kumkal}], $\langle n(\Delta t) \rangle$ is the mean number of electrons flowing through the molecule in the time interval $\Delta t$ and $\langle \Delta n([\Delta t)]^2 \rangle$ is the variance of this number. It is equivalent to the Fano factor defined as $F=S(0)/2e\langle I \rangle$, where $S(0)$ is the zero-frequency noise power and $\langle I \rangle$ is the mean current~\cite{nazarov2009}. In the considered system the Fano factor reads
\begin{align}  \label{fanoexpr}
F=\frac{\Gamma_L^2 \left(1-p_L^2 p_R^2-2 p_L p_R+2 p_R^2 \right)+\Gamma_R^2 \left(p_R^2-1\right)^2}{\left[\Gamma_L (p_L p_R-1)+\Gamma_R \left(p_R^2-1\right)\right]^2}.
\end{align}
The noise in the system can be either sub-Poissonian ($F<1$) or super-Poissonian ($F>1$). The former case is characteristic for systems with a Coulomb blockade~\cite{bagrets2003}. For $p_L=p_R=0$ and $\Gamma_L=\Gamma_R$ the Fano factor reaches the minimal value 1/2 -- the same as in the symmetrically coupled single quantum dot in the strong Coulomb blockade regime~\cite{bagrets2003}. On the other hand, the super-Poissonian noise enhancement, appearing for sufficiently large values of $p_R$, at first look can be interpreted as a result of the phenomenon referred to as the dynamical channel blockade~\cite{bulka2000, belzig2005}. This phenomenon can be described as follows: competition between transport of $\uparrow$ and $\downarrow$ electrons, corresponding to either the fast or the slow tunneling to the right lead, increases the randomness of the number of particles transferred in a given time interval, thus leading to the noise enhancement. As a matter of fact, the expression for the Fano factor given by Eq.~\eqref{fanoexpr} is exactly the same as in the case of single quantum dot in a strong Coulomb blockade regime attached to spin-polarized leads, the exemplary case of the dynamical channel blockade~\cite{bulka2000}. One should be aware, however, that the dynamical channel blockade in itself do not generate the waiting time correlations~\cite{ptaszynski2017}, which -- as it will be shown -- are present in the considered system.

It is also apparent, that the Fano factor depends on the parameters $\Gamma_L$, $\Gamma_R$, $p_L$ and $p_R$, however is independent of $J$. The same was found to be true also for higher cumulants of FCS. This contrasts with the dependence of the variance and higher cumulants of WTD on $J$. To be more explicit, let us compare the Fano factor and the previously considered randomness parameter $R_{LR}$. I focus on the case of $\Gamma_L \gg \Gamma_R$, when the dynamics of the system is wholly determined by the slowest processes -- tunneling through the right lead and coherent oscillations between spin states. Such a choice of parameters is very convenient since it does not affect the most interesting phenomena taking place in the system, because the tunneling through the left lead is a trivial Poissonian process; on the other hand, it greatly simplifies the interpretation of results by removing the uninteresting dependence of the studied quantities on $\Gamma_L$. In such a limit, for $p_L=0$, the Fano factor is given by the very simple expression \begin{equation} \label{fanosimpl}
F=1+2p_R^2.
\end{equation}
In systems described by the renewal theory, which assumes that the successive waiting times are independent and identically distributed random variables (the assumption referred to as the renewal property), the identity $F=R_{LL}=R_{RR}$ holds~\cite{albert2011, budini2011}. Therefore, in the considered limit of $\Gamma_L \gg \Gamma_R$ (when $R_{LL}=R_{RR}=R_{LR}$), the identity $F = R_{LR}$ should hold if the system is renewal. A comparison of Eqs.~\eqref{rlrpl0} and~\eqref{fanosimpl} shows that this identity is satisfied only for $J=0$ [cf. Fig.~\ref{fig:fanorandkoh}], $p_R =0$ (when the tunneling to the right lead is spin-independent) or $p_R=\pm 1$ (when the current is blocked). This indicates that for $|J|>0$ with $|p_R| \in (0,1)$ the renewal property does not hold any longer, which means that the subsequent waiting times are correlated~\cite{albert2012}.

Why the zero-frequency FCS does not depend on $J$, while WTD does? It results from the fact, that the zero-frequency FCS is sensitive only to the long-time behavior, while WTD focuses on the short-time one. It appears, that the long-time behavior is determined by the values of tunneling rates for different spin polarizations, as well as by populations of electrons with specific spin orientations in the right dot. Both factors are not influenced by the oscillations between the spin states because, as previously mentioned, the coherent oscillations leave the spin population of the right dot intact. One should note, however, that this lack of the dependence of FCS on $J$ is a feature of the specific model considered in this paper, which assumes the unidirectionality of transport. When the assumption that all relevant system states are well within the transport window does not hold any longer and thermally excited tunneling cannot be neglected, $J$ appears to affect both the current and the zero-frequency FCS~\cite{xue2014, golovach2004, michalek2006, michalek2011}. However, even is such a case its influence is often not very significant~\cite{weymann2008}. It should be also noted that the dependence on $J$ can be observed when one analyzes the finite-frequency FCS. However, it does not provide much new information in comparison with WTD, therefore I do not discuss this issue further in this paper.

\begin{figure} 
	\centering
	\includegraphics[width=0.9\linewidth]{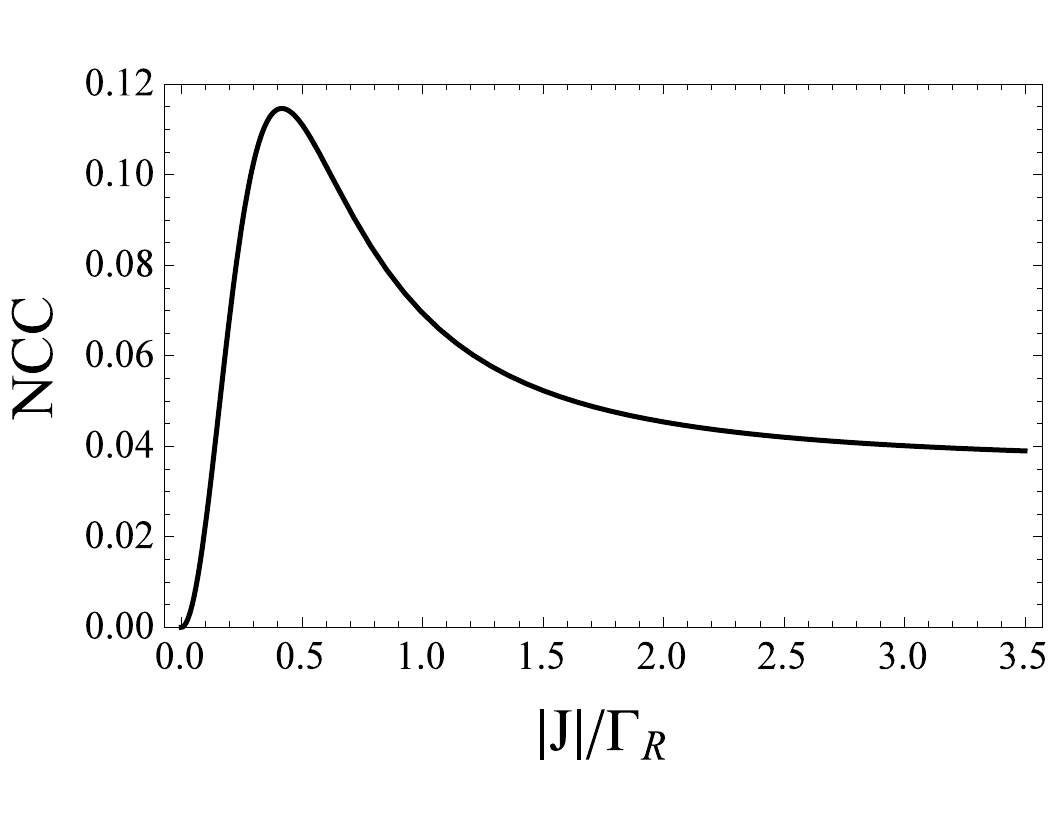} 
	\caption{Normalized cross-correlation of two successive times $\tau_{RR}$ for $p_L=0$, $p_R=\sqrt{2}/2 \approx 0.7$ and $\Gamma_L \gg \Gamma_R$.}
	\label{fig:ncckoh}
\end{figure}

As mentioned, the inequality of the Fano factor and the randomness parameter implies the breaking of the renewal property, which can be shown directly by determining the correlation of the subsequent waiting times~\cite{budini2010, dasenbrook2015, ptaszynski2017}. Figure~\ref{fig:ncckoh} shows the dependence of the normalized cross-correlation of waiting times $\tau_{RR}$ [see Eq.~\eqref{ncc}] on $|J|/\Gamma_R$. As one can observe, for $J=0$ the correlation does not occur. As a matter of fact, in this case the joint distribution of two subsequent waiting times $\tau_{RR}$, denoted as $w_{RR}(\tau_1,\tau_2)$, is equal to the product of two single waiting time distributions: $w_{RR}(\tau_1,\tau_2)=w_{RR}(\tau_1)w_{RR}(\tau_2)$. This results from the absence of interaction between spins, which makes tunneling events uncorrelated. 

Then, I consider the case of $|J| \gg \Gamma_R$, focusing on the case of $p_L=0$. The dynamics is described by the effective five-state model [Fig.~\ref{fig:6statescoh}~(b)].
Let us analyze the process starting with the tunneling from the ${|\! \! \uparrow \uparrow \rangle}$ state. For $p_R>0$, the waiting time $\tau_{LR}$ associated with this transition would be probably shorter than the mean. Then, due to $p_L=0$, the state ${|\! \! \uparrow \uparrow \rangle}$ or the subspace ${|S_z=0 \rangle}$ would be generated with the equal probability. Therefore, the next waiting time would be probably either shorter than the mean one if the state ${|\! \! \uparrow \uparrow \rangle}$ was generated, or close to the mean one if the subspace ${|S_z=0 \rangle}$ was produced. Thus, the subsequent waiting times would tend to be positively correlated. The positive waiting time correlations in the process starting from the tunneling ${|\! \! \downarrow \downarrow \rangle}$ can be shown in a similar way. In result, the whole dynamics is associated with the positive cross-correlation of the successive waiting times.

Lastly, NCC reaches its maximum for some finite value of $|J|/\Gamma_R$. In such a case, the mechanism leading to the cross-correlation is much more complex. Let us consider a one factor which seams to be important. For $p_R>0$, if the first waiting time $\tau_{RR}$ is long, it can be supposed that either the ${|\! \! \downarrow \downarrow \rangle}$ or ${|\! \! \uparrow \downarrow \rangle}$ state was previously generated. If the latter is the case, with the high probability the electron jump occurred only after the rotation to the ${|\! \! \downarrow \uparrow \rangle}$ state took place. Thus, whichever of states ${|\! \! \downarrow \downarrow \rangle}$ or ${|\! \! \uparrow \downarrow \rangle}$ was previously generated, the jump to right lead would leave the molecule in the ${|\! \downarrow \rangle}$ rather than ${|\! \uparrow \rangle}$ subspace. This, in turn, leads to the increased probability that the next waiting time will be also longer then the mean one, which results in the positive waiting time cross-correlation.

\subsection{\label{subsec:gbpos}$\nabla B \neq 0$}
The previous section has shown that the waiting time distribution is dependent on the value of the exchange coupling $J$, and therefore on the frequency of the coherent oscillations. However, the dependence of the randomness parameter and the cross-correlation of waiting times is visible only for small values of $J$ -- for sufficiently high values of $J$ (more or less $|J|>3 \Gamma_R$) they become nearly constant. In experimentally relevant situations, however, the exchange coupling may be much larger than the values of the tunneling rates, which in single electron counting experiments must be sufficiently small due to limited frequency bandwidth of detectors~\cite{gustavsson2009}. This section shows that the dependence of the waiting time distribution on the spin dynamics can be observable also for high values of $J$ when one applies the inhomogeneous magnetic field ($\nabla B \neq 0$), which changes the direction of the qubit rotation axis.   
 %
 \begin{figure} 
 	\centering
 	\includegraphics[width=0.9\linewidth]{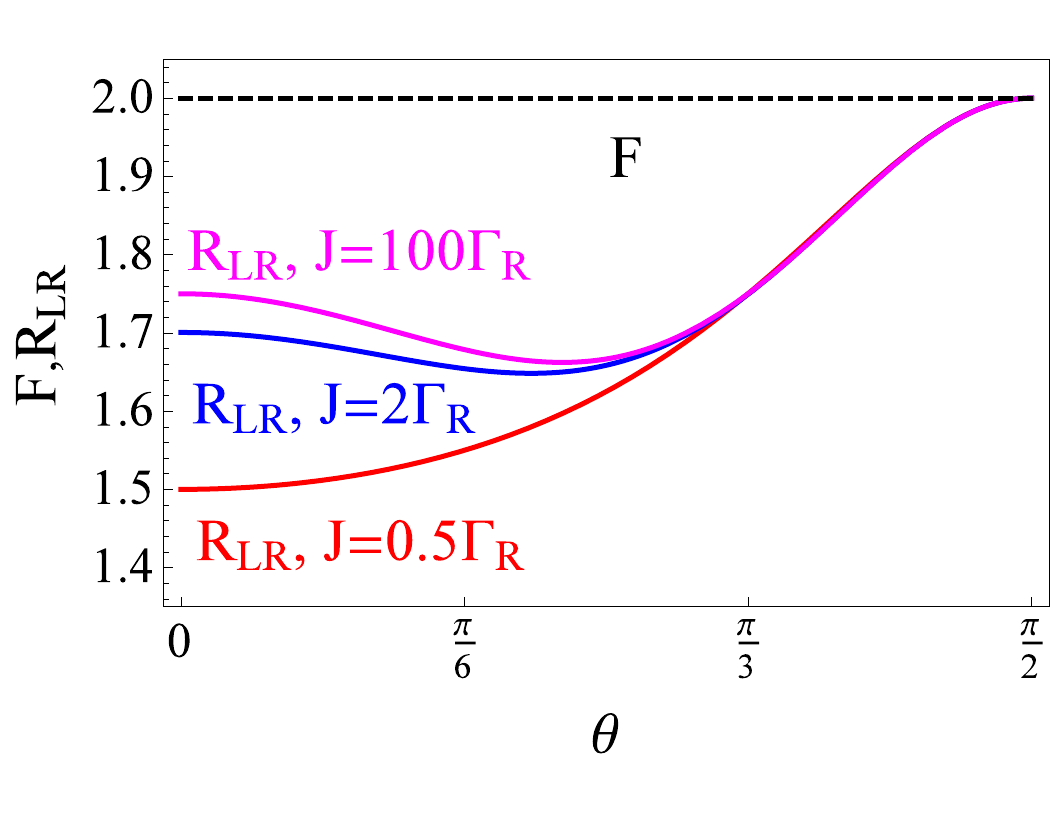} 
 	\caption{Dependence of the Fano factor $F$ and the randomness parameter $R_{LR}$ on the angle between z-axis of the Bloch sphere and the rotation axis denoted as $\theta$. Black dashed line -- the Fano factor (the same for all values of $J$ and $\theta$), red line -- $J=0.5 \Gamma_R$, blue line -- $J=1.5 \Gamma_R$, violet line -- $J=100 \Gamma_R$. All results for $p_L=0$ and $p_R=\sqrt{2}/2 \approx 0.7$. Fano factor calculated in the limit of $\Gamma_L \gg \Gamma_R$.}
 	\label{fig:fanorandkoodthh}
 \end{figure}
 %

As in the previously considered case, the assumption $|t_{12}| \gg |J|, \Gamma_L, \Gamma_R$ is taken. It is also assumed that $|t_{12}| \gg \gamma |\Delta B|$. This assumption provides, that in both singly-occupied states probabilities of the occupancy of the first and the second dot are the same and therefore, as in the previous subsection, one may work in the reduced space [with the Liouvillian presented in Eq.~\eqref{liovred} in the Appendix]. Figure~\ref{fig:fanorandkoodthh} shows the dependence of the Fano factor $F$ and the randomness parameter $R_{LR}$ on the angle between z-axis of the Bloch sphere and the qubit rotation axis denoted as $\theta$ (see Fig.~\ref{fig:bloch}). Similarly to the previously considered case, the Fano factor is given by Eq.~\eqref{fanoexpr} and does not depend on either $J$ or $\theta$. The randomness parameter is, on the other hand, dependent on both parameters. The dependence on the angle $\theta$ is observable also for high values of $J$. Interestingly, the value of the exchange coupling $J$ affects the character of the dependence on the angle $\theta$ qualitatively. For small values of $J$ the randomness parameter rises monotonically in the range of $\theta \in [0, \pi/2]$. It can be interpreted as the result of the reduced probability of the ${|\! \! \uparrow \downarrow \rangle} \leftrightarrow {|\! \! \downarrow \uparrow \rangle}$ transition due to the increased value of $\Delta B$. For $\theta=\pi/2$ the oscillations between these states do not occur, which corresponds to the case of $J=0$ from the previous section; in result the randomness parameter $R_{LR}$ reaches its maximum and, for $\Gamma_L \gg \Gamma_R$, is equal to the Fano factor.

For high values of $J$ the dependence on $\theta$ is, on the other hand, non-monotonic. To explain this behavior qualitatively, I will focus on the limit of $|J| \gg \Gamma_R$ and consider dynamics of the system described by the diagonalized master equation (DME)~\cite{poltl2009}. In contrast to GME, this approach uses the basis of the eigenstates of the Hamiltonian $\hat{H}_D$ instead of the basis of the localized states, and therefore only the dynamics of diagonal elements of the density matrix (state probabilities) is considered. In general, DME does not fully describe the coherent dynamics of the system~\cite{poltl2009}. However, it appears that for high values of $J$ it captures the dependence on the angle $\theta$ quite well, because the eigenstates are modified by the presence of the magnetic field gradient and, on the other hand, timescales of the coherent oscillations and electron tunneling are well separated.

%
\begin{figure} 
	\centering
	\includegraphics[width=0.75\linewidth]{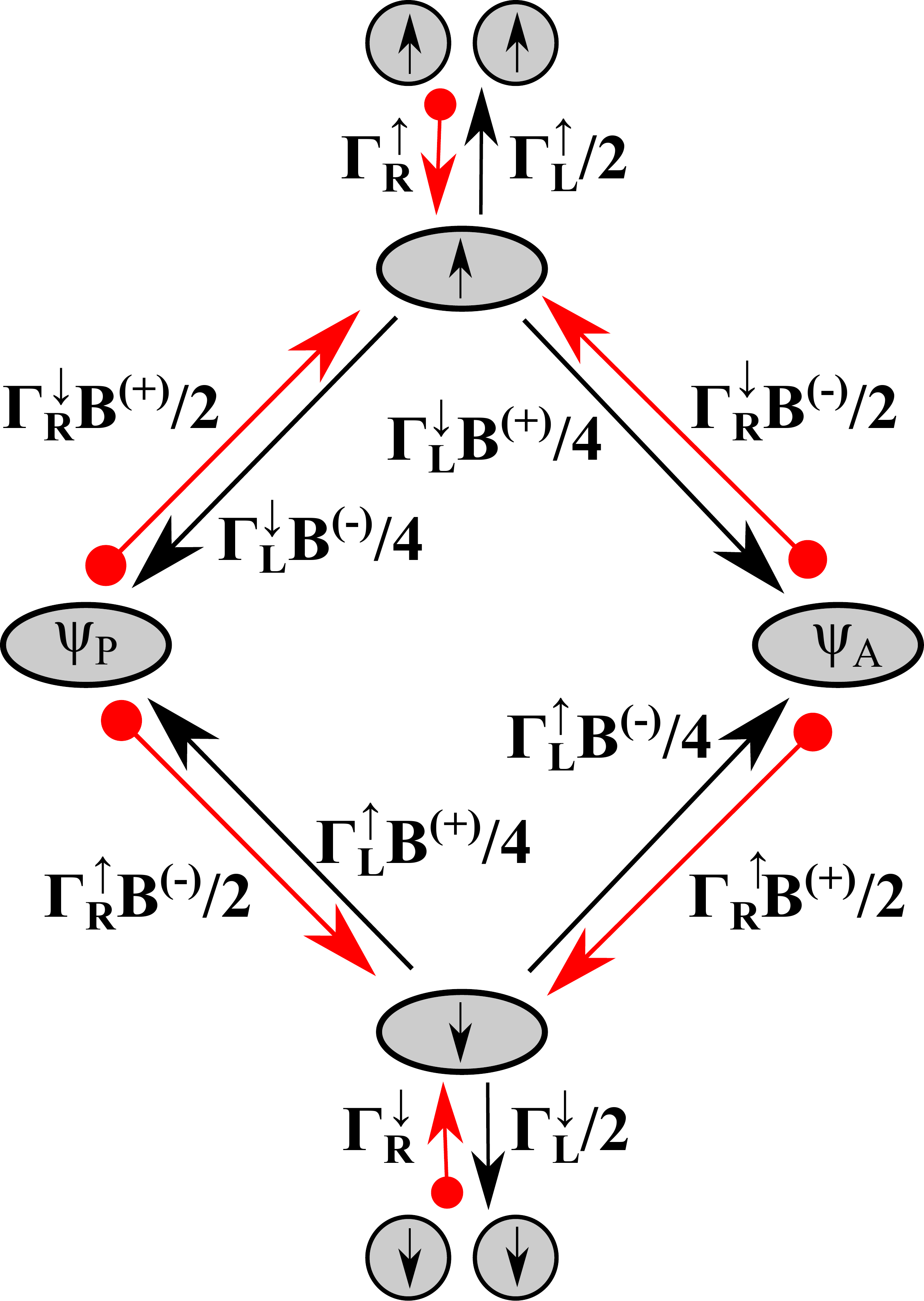} 
	\caption{Six-state model of the dynamics of the system for $\nabla B \neq 0$ and $J \gg \Gamma_R, \Gamma_L$. Red arrows with bullet tails correspond to the tunneling from the molecule to the right lead, while black arrows to the tunneling from the left lead to the molecule. $\Gamma_{\alpha}^{\uparrow}=(1+p_{\alpha})\Gamma_{\alpha}$, $\Gamma_{\alpha}^{\downarrow}=(1-p_{\alpha})\Gamma_{\alpha}$, $B^{(\pm)}=1 \pm \sin \theta$. States $|\psi_P \rangle$ and $|\psi_A \rangle$ as in Eqs.~\eqref{psi1}--\eqref{psi2}.}
	\label{fig:6stanowdme}
\end{figure}
%

In the DME approach, the rate equation for the single diagonal element $\rho_{ii}$ (the probability of the eigenstate $|\phi_i \rangle$) reads
\begin{align} \label{dme}
\frac{d \rho_{{ii}}}{dt} = \sum_{j \neq i} \left(\Gamma_{j \rightarrow i} \rho_{jj} - \Gamma_{i \rightarrow j} \rho_{ii} \right),
\end{align}
where
\begin{align} \label{rate}
\Gamma_{i \rightarrow j}&=\sum_{\sigma \sigma'} \Gamma_L^{\sigma}  \left|\langle \phi_j |l^{\dagger}_{\sigma \sigma'}|\phi_i \rangle\right|^2 \\ \nonumber &+ \sum_{\sigma \sigma'} \Gamma_R^{\sigma'}  \left|\langle \phi_j |r^{\dagger}_{\sigma \sigma'}|\phi_i \rangle \right|^2 .
\end{align}
and the operators $l^{\dagger}_{\sigma \sigma'}$ and $r^{\dagger}_{\sigma \sigma'}$ are defined as in Eq.~\eqref{lindblad}. Using these rate equations one obtains the effective six-state model (Fig.~\ref{fig:6stanowdme}) in the basis $\{ {|\! \uparrow \rangle }, {|\! \downarrow \rangle }, {|\! \! \uparrow \uparrow \rangle}, {| \psi_P \rangle}, {|\psi_A \rangle}, {|\! \! \downarrow \downarrow \rangle}\}$, where the ${| \psi_P \rangle}$ and ${|\psi_A \rangle}$ states were defined in Eqs.~\eqref{psi1}--\eqref{psi2}. The Liouvillian corresponding to this model is presented in Eq.~\eqref{liovdme} in the Appendix. It can be noted that the tunneling rates are linear functions of $\sin \theta$ (see the description of Fig.~\ref{fig:6stanowdme}). Therefore, in the following I will focus on the dependence of the analyzed quantities on $\sin \theta$ rather than on the angle $\theta$ itself.

%
\begin{figure} 
	\centering
	\includegraphics[width=0.9\linewidth]{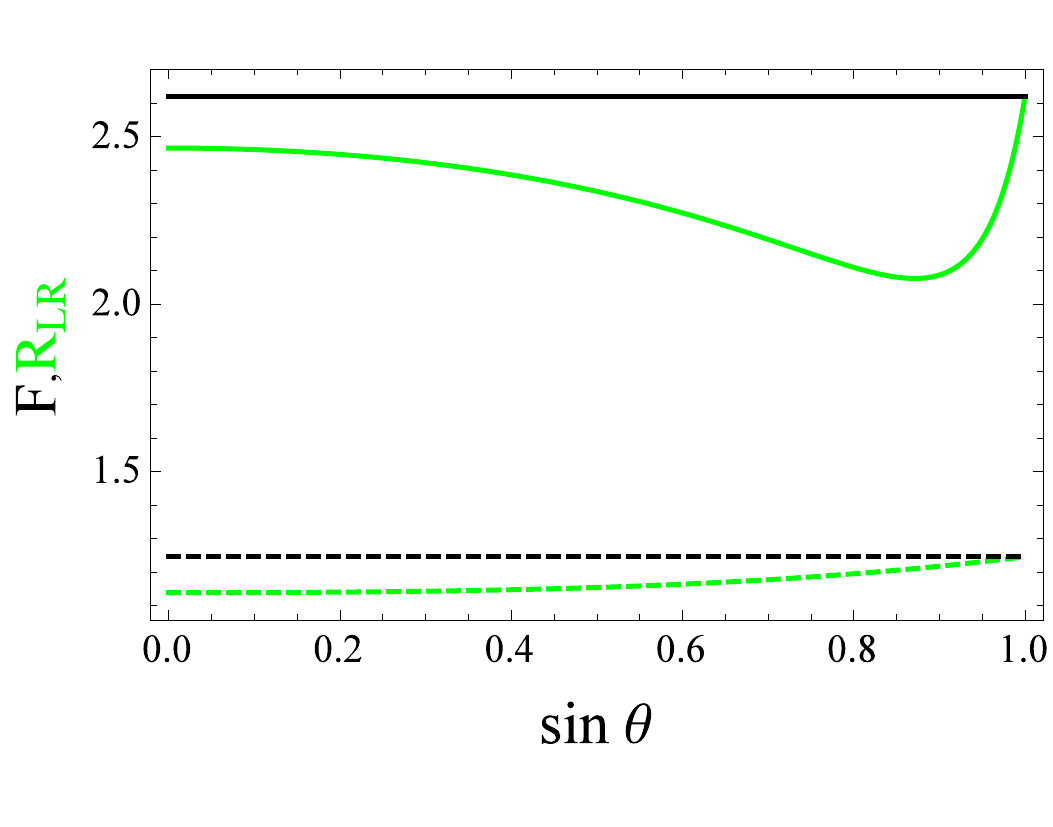} 
	\caption{Dependence of the Fano factor $F$ (black lines) and the randomness parameter $R_{LR}$ (green lines) on $\sin \theta$ for $p_L=0$ and $p_R=0.35$ (dashed lines) or $p_R=0.9$ (solid lines) in the limit of $J \gg \Gamma_L, \Gamma_R$. Fano factor calculated in the limit of $\Gamma_L \gg \Gamma_R$.}
	\label{fig:fanoranddme}
\end{figure}
%

\begin{figure} 
	\centering
	\includegraphics[width=0.9\linewidth]{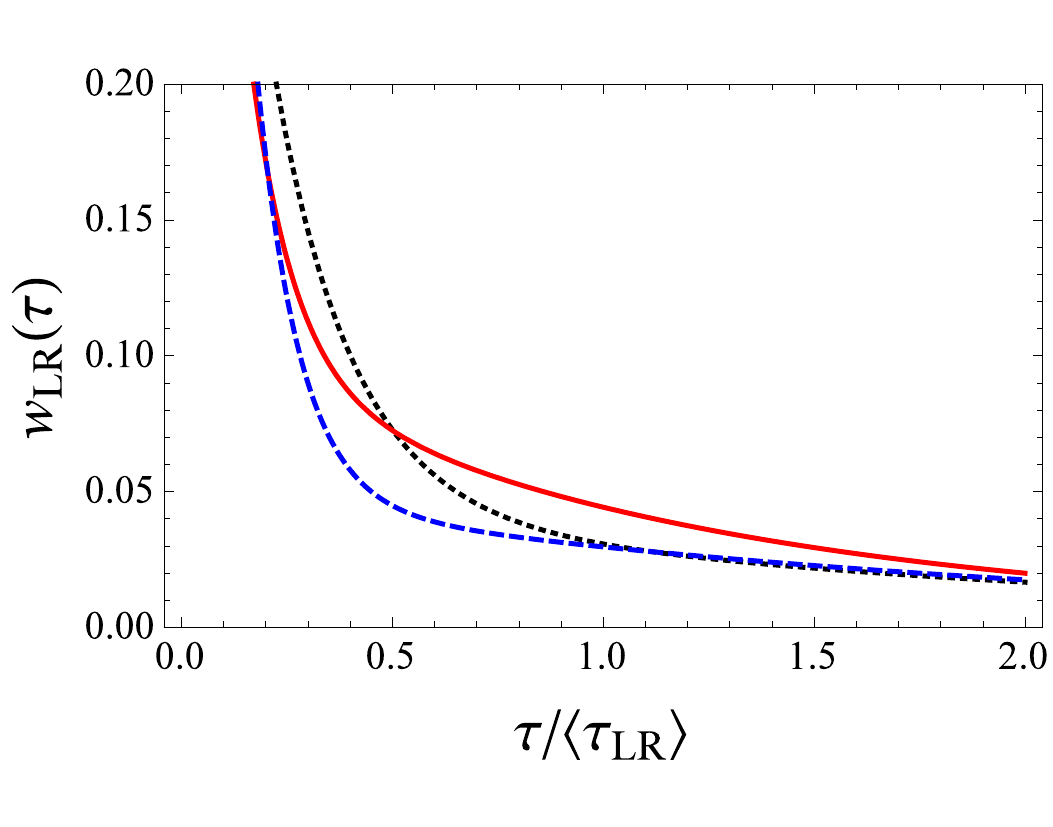} 
	\caption{Waiting time distribution $w_{LR}(\tau)$ for $\sin \theta=0$ (black dotted line), $\sin \theta=0.875$, which is close to the minimum of $R_{LR}$ (red solid line) and $\sin \theta =1$ (blue dashed line) in the limit of $J \gg \Gamma_L, \Gamma_R$. Results for $p_L=0$ and $p_R=0.9$.}
	\label{fig:wtddme}
\end{figure}

Figure~\ref{fig:fanoranddme} shows the dependence of the Fano factor and the randomness parameter $R_{LR}$ on  $\sin \theta$ with $p_L=0$ and  $\Gamma_L \gg \Gamma_R$ for two different values of $p_R$. One can observe, that the shape of $R_{LR}-\sin \theta$ curve depends on the value of $p_R$. For relatively small $p_R=0.35$ the dependence of the randomness parameter on $\sin \theta$ is monotonic, similarly to the case of small $J$ in Fig.~\ref{fig:fanorandkoodthh}. As previously, it can be explained as the result of the reduced probability of the ${|\! \! \uparrow \downarrow \rangle} \leftrightarrow {|\! \! \downarrow \uparrow \rangle}$ transition, here defined as $\langle \psi_i |\! \! \uparrow \downarrow \rangle \langle \downarrow \uparrow  \! \! | \psi_i \rangle$ (with $i \in \{A,P\}$). The non-monotonic behavior appears, on the other hand, for high $p_R=0.9$. As Fig.~\ref{fig:wtddme} shows, the reduction of the randomness parameter for $\sin \theta \approx 0.875$ is the result of the increased probability of the waiting time being close to the mean. To explain this behavior of the waiting time distribution $w_{LR}(\tau)$, let us analyze its exact form obtained using the DME approach:
\begin{align} \label{wbfin}
w_{LR}(\tau)&= A_{\uparrow \uparrow} e^{-(1+p_R) \Gamma_R} + A_{\psi_P} e^{-(1- p_R \sin \theta) \Gamma_R} \\ \nonumber &+ A_{\psi_A} e^{-(1+ p_R \sin \theta) \Gamma_R} +A_{\downarrow \downarrow} e^{-(1-p_R) \Gamma_R}.
\end{align}
%
%
%
\begin{figure} 
	\centering
	\includegraphics[width=0.9\linewidth]{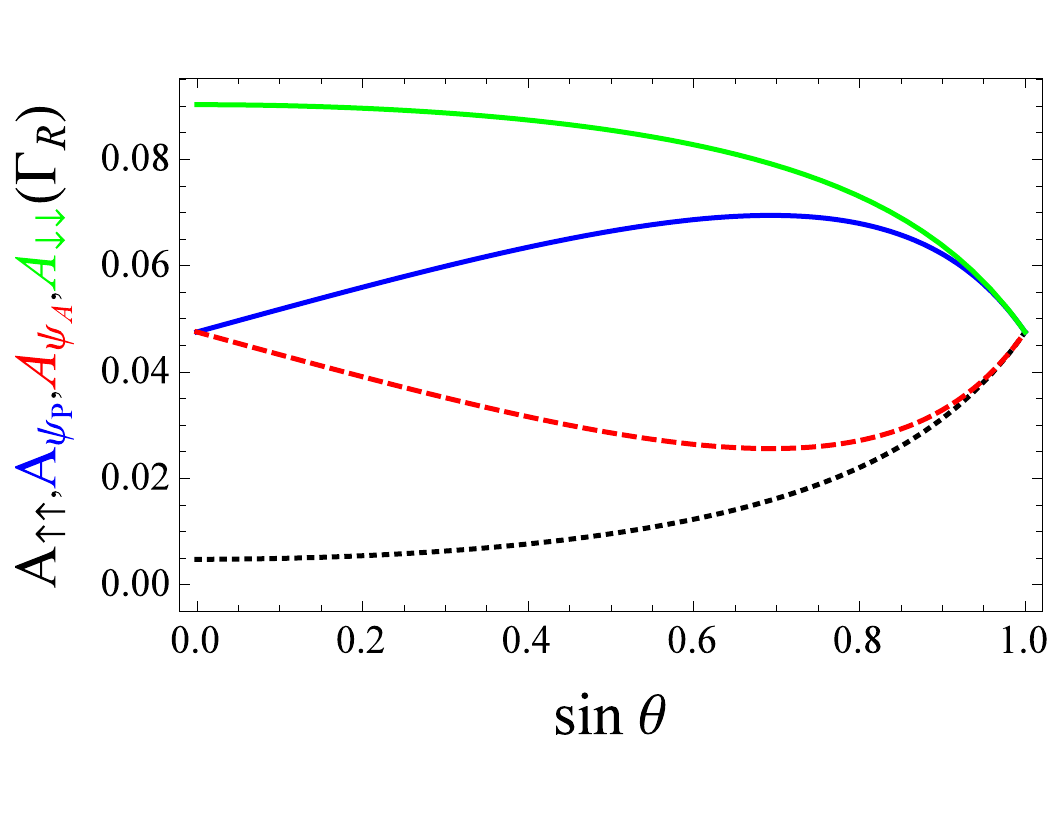} 
	\caption{Dependence of amplitudes of terms of Eq.~\eqref{wbfin} on $\sin \theta$ in the limit of $J \gg \Gamma_L, \Gamma_R$ for $p_L=0$ and $p_R=0.9$. $A_{\uparrow \uparrow}$ -- black dotted line, $A_{\psi_P}$ -- blue solid line, $A_{\psi_A}$ -- red dashed line, $A_{\downarrow \downarrow}$ -- green solid line.}
	\label{fig:amplitudes}
\end{figure}
%
The successive terms of this formula are the waiting time distributions corresponding to the tunneling from the states ${|\! \! \uparrow \uparrow \rangle}, {| \psi_P \rangle}, {|\psi_A \rangle}$ and ${|\! \! \downarrow \downarrow \rangle}$, respectively. For $\sin \theta=0$ and $\sin \theta =1$ the dynamics is analogous to the cases of $|J| \gg \Gamma_R$ and $J=0$, respectively, and therefore the formula is reduced to the triple-exponential or the double-exponential form given by Eq.~\eqref{wjinf} or Eq.~\eqref{wj0}. Now, let us analyze the behavior of the amplitudes $A_{\uparrow \uparrow}$, $A_{\psi_A}$, $A_{\psi_P}$, $A_{\downarrow \downarrow}$, which are equal to probabilities that the states ${|\! \! \uparrow \uparrow \rangle}, {| \psi_A \rangle}, {|\psi_P \rangle}$ or ${|\! \! \downarrow \downarrow \rangle}$ are generated, multiplied by the corresponding departure rates from these states. I focus on the case of $\sin \theta>0$ and $p_R=0.9$. Results are shown in Fig.~\ref{fig:amplitudes}. For $\sin \theta=0$ the amplitude $A_{\uparrow \uparrow}$ is significantly low. This can be explained as follows: due to $p_R \approx 1$ transitions from the ${| \psi_P \rangle}$ and ${|\psi_A \rangle}$ states to the ${|\! \uparrow \rangle }$ subspace are suppressed, and the dynamics of the systems is dominated by the transitions occurring in the lower part of Fig.~\ref{fig:6stanowdme} (containing the states ${| \psi_P \rangle}$,${|\psi_A \rangle}$, ${|\! \! \downarrow \downarrow \rangle}$ and the ${|\! \downarrow \rangle }$ subspace). Probabilities of the subspace ${|\! \uparrow \rangle }$ and the state ${|\! \! \uparrow \uparrow \rangle}$, and consequently also the amplitude $A_{\uparrow \uparrow}$, are significantly reduced. When $\sin \theta$ rises, the amplitude $A_{\psi_P}$ increases, and it exceeds the amplitude $A_{\psi_A}$ because, due to the high value of $\sin \theta$, the transitions from the ${|\! \downarrow \rangle }$ subspace to the ${| \psi_A \rangle}$ state are suppressed. As one may notice, the increased probability of the waiting time being close to the mean for $\sin \theta=0.875$ (Fig.~\ref{fig:wtddme}, red solid line), and therefore also the reduced value of the randomness parameter, is associated with the large value of the amplitude $A_{\psi_P}$. As a matter of fact, the departure rate from the $ {| \psi_P \rangle}$ state, equal to $1/(1- p_R \sin \theta)\Gamma_R$, is very close to the mean waiting time $1/(1-p_R^2)\Gamma_R $, which makes WTD more concentrated around the mean value. However, at the same time, for high values of $\sin \theta$ the amplitude $A_{\uparrow \uparrow}$ also rises due to the increased rate of generation of the subspace $ {|\! \uparrow \rangle }$ by tunneling from the state $ {| \psi_P \rangle}$. This, alongside with the blocking of the $ {|\! \uparrow \rangle } \rightarrow {| \psi_P \rangle}$ transition, leads to the decrease of the amplitude $A_{\psi_P}$ for $\sin \theta$ close to 1, which increases the value of the randomness parameter. Finally, for $\sin \theta=1$ all amplitudes become equal, since the oscillations between the spin states do not occur, and therefore all states are generated with the same probability.

%
\begin{figure} 
	\centering
	\includegraphics[width=0.9\linewidth]{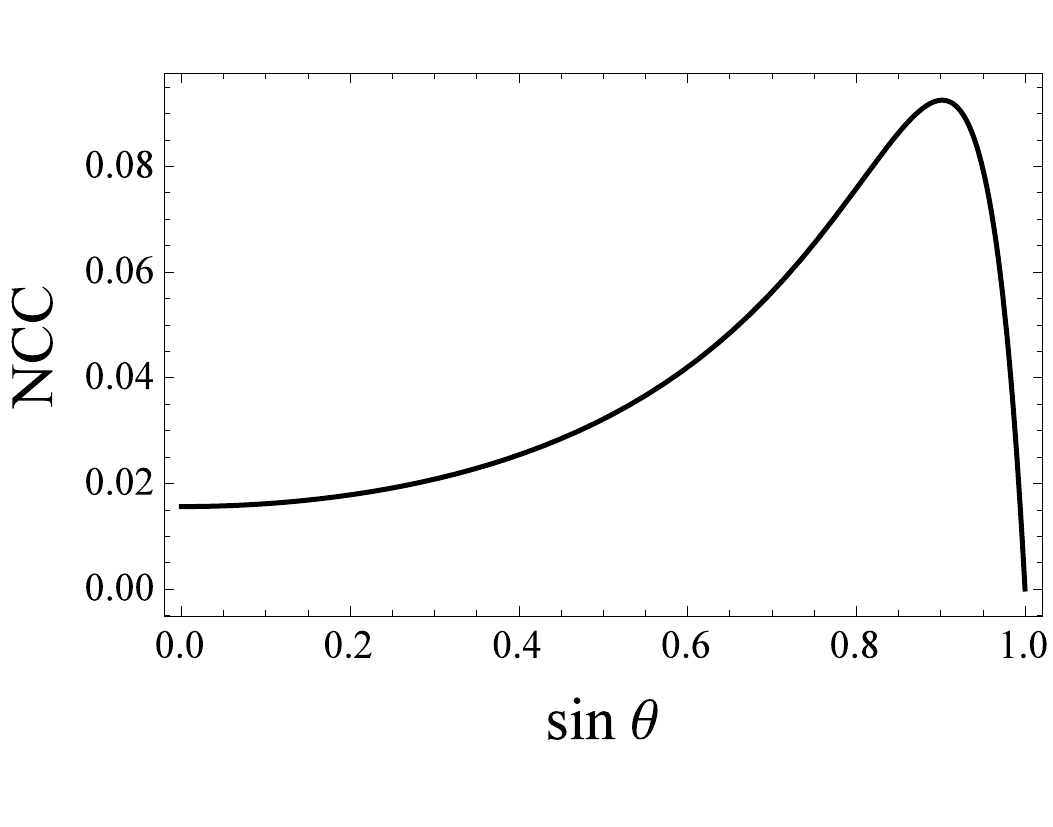} 
	\caption{Normalized cross-correlation of two successive times $\tau_{RR}$ in the limit of $J \gg \Gamma_L, \Gamma_R$ for $p_L=0$, $p_R=0.9$ and $\Gamma_L \gg \Gamma_R$.}
	\label{fig:nccdme}
\end{figure}
%

As in the case of $\nabla B=0$, the inequality of the randomness parameter and the Fano factor is the result of the breaking of the renewal property, which is well revealed by the presence of the waiting time correlations. As Fig.~\ref{fig:nccdme} indicates, for $p_L=0$ and $p_R=0.9$ the normalized cross-correlation exhibits the non-monotonic dependence on $\sin \theta$, which corresponds to the non-monotonic behavior of the randomness parameter. Since the limits of $\sin \theta=0$ and $\sin \theta=1$ correspond to the cases of $|J| \gg \Gamma_R$ and $J=0$ from Sec.~\ref{subsec:gb0}, let us focus on the case of $\sin \theta \approx 0.875$, when the cross-correlation reaches the maximum. As can be inferred from a model of the dynamics shown in Fig.~\ref{fig:6stanowdme}, for large values of $\sin \theta$ transitions ${|\! \downarrow \rangle } \rightarrow {|\psi_A \rangle}$ and ${|\! \uparrow \rangle } \rightarrow {|\psi_P \rangle}$ are suppressed. Thus, the dynamics of the system can be described as follows: the initial state is one of the states or subspaces of the top right part of Fig.~\ref{fig:6stanowdme} (${|\! \! \uparrow \uparrow \rangle}$, $ {|\! \uparrow \rangle}$ or ${| \psi_A \rangle}$). After the certain time spent in the top right part, transition $ {|\psi_A \rangle} \rightarrow {|\! \downarrow \rangle }$ takes place, and then the evolution occurs within the bottom left part of Fig.~\ref{fig:6stanowdme} (containing the ${|\! \! \downarrow \downarrow \rangle}$, ${| \psi_P \rangle}$ states and the $ {|\! \downarrow \rangle}$ subspace), until the transition $ {|\psi_P \rangle} \rightarrow {|\! \uparrow \rangle }$ is realized. Thus, one can observe the switching between the two subspaces of the whole system, which are associated with different timescales of the tunneling to the right lead: the states ${|\! \! \uparrow \uparrow \rangle}$ and ${| \psi_A \rangle}$ from the top right part are associated with the high departure rates [$(1+p_R) \Gamma_R$ and $(1+ p_R \sin \theta) \Gamma_R$, respectively], while the states ${|\! \! \downarrow \uparrow \rangle}$ and ${| \psi_P \rangle}$ from the bottom left part are associated with the low departure rates [$(1-p_R) \Gamma_R$ and $(1- p_R \sin \theta) \Gamma_R$, respectively]. The stochastic switching between the transport channels with different timescales of the tunneling, referred to as the telegraphic switching, has been already shown to generate the waiting time correlations~\cite{ptaszynski2017}. For $\sin \theta$ close to 1, however, the other process starts to be important: transport in the clockwise direction along the trajectory ${| \psi_P \rangle} \rightarrow {|\! \uparrow \rangle } \rightarrow  {| \psi_A \rangle} \rightarrow {|\! \downarrow \rangle } \rightarrow {| \psi_P \rangle}$. It leads to the intermingling of short and long waiting times, which reduces the value of the cross-correlation, and finally makes it vanish for $\sin \theta=1$.

\section{Conclusions} \label{sec:conclusions}
Unidirectional electron transport in the double quantum dot attached to the spin-polarized leads has been studied by means of the Markovian quantum master equation. The study has focused on the analysis of current fluctuations described by the zero-frequency full counting statistics and the waiting time distribution. In particular, the dependence of FCS and WTD on the exchange coupling between dots $J$ and the difference of values of the magnetic field applied to dots $\Delta B$ has been investigated. The waiting time distribution has been shown to exhibit a quite nontrivial dependence on $J$ and $\Delta B$, well revealed by the non-monotonic behavior of the randomness parameter. It is a result of the rich and complex internal dynamics associated with the coherent oscillations between the spin states. In contrast, the zero-frequency full counting statistics is in no way dependent on these parameters. It only reveals the noise enhancement resulting from the spin polarization of leads, and does not provide any information about the internal dynamics of the molecule. Importantly, such a nonequivalence of the information provided by the used approaches is possible only when the subsequent waiting times are correlated and therefore the renewal theory, relating FCS and WTD, is not applicable. As a matter of fact, breaking of the renewal property is clearly shown by determining the cross-correlation of the subsequent waiting times.

These results lead to the more general conclusion: there are transport systems, for which the waiting time distribution gives information about the hidden dynamics of the system which is not provided by the zero-frequency full counting statistics. It can be explained by the sensitivity of WTD to the short-time dynamics, which contrasts with the focusing of the zero-frequency FCS on the long-time one. This highlights the usefulness of the analysis of the waiting time distribution in the field of electronic transport, as well as in the study of the internal dynamics of other types of open quantum systems, for example optical ones. Moreover, although the effects analyzed in this paper are quantum mechanical in nature, the conclusion drawn here is not restricted to quantum systems, since Markovian models similar to the one shown in Fig.~\ref{fig:6stanowdme} can describe also the classical stochastic dynamics. This fact can be important, for example, for the developing field of statistical kinetics of biomolecular systems~\cite{chemla2008, moffitt2010, barato2015, shaevitz2005}.

\section*{Acknowledgments}
I thank B. R. Bu\l{}ka for the careful reading of the manuscript and the valuable discussion, and J. Martinek for useful remarks. This work has been supported by the National Science Centre, Poland, under the project 2016/21/B/ST3/02160.  

\appendix*

\section{\label{sec:appendix}Matrix form of the Liouvillian}
Here I present the full matrix form of the Liouvillian obtained using Eq.~\ref{lindblad}. The following notation of the system states is used: ${|1 \rangle} \equiv {|\! \! \uparrow \! \!  0 \rangle}$, ${|2 \rangle} \equiv {|\! \!  \downarrow \! \! 0 \rangle}$, ${|3 \rangle} \equiv {|0 \! \! \uparrow \rangle}$, ${|4 \rangle} \equiv {|0 \! \! \downarrow \rangle}$, ${|5 \rangle} \equiv {|\! \! \uparrow \uparrow \rangle}$, ${|6 \rangle} \equiv {|\! \! \uparrow \downarrow \rangle}$, ${|7 \rangle} \equiv {|\! \! \downarrow \uparrow \rangle}$, ${|8 \rangle} \equiv {|\! \! \downarrow \downarrow \rangle}$. The state vector in the Liouville space is defined as
\begin{equation} \label{fullvec} \rho=(\rho_{11},\rho_{22},...,\rho_{88},\mathcal{R}_{13},\mathcal{I}_{13},\mathcal{R}_{24}, \mathcal{I}_{24}, \mathcal{R}_{67}, \mathcal{I}_{67})^T,
\end{equation} 
where $\rho_{ij}$ are elements of the density matrix, $\mathcal{R}_{ij}=\Re[\rho_{ij}]$ and $\mathcal{I}_{ij}=\Im[\rho_{ij}]$. All coherences (off-diagonal elements of the density matrix) not included within the defined vector can be neglected, because they are not generated by any term of the Liouvillian. The Liouvillian generating the evolution of the state vector takes the form
\begin{widetext}
\setcounter{MaxMatrixCols}{14}	
\begin{equation} \label{fullliov} \mathcal{L}=
\begin{pmatrix}
0 & 0 & 0 & 0 & \mathbf{\Gamma_R^\uparrow} & \mathbf{\Gamma_R^\downarrow} & 0 & 0 & 0 & -2t_{12} & 0 & 0 & 0 & 0\\
0 & 0 & 0 & 0 & 0 & 0 & \mathbf{\Gamma_R^\uparrow} & \mathbf{\Gamma_R^\downarrow} & 0 & 0 & 0 & -2 t_{12} & 0 & 0 \\
0 & 0 & -2 \Gamma_L & 0 & 0 & 0 & 0 & 0 & 0 & 2t_{12} & 0 & 0 & 0 & 0 \\
0 & 0 & 0 & -2 \Gamma_L & 0 & 0 & 0 & 0 & 0 & 0 & 0 & 2t_{12} & 0 & 0 \\
0 & 0 & \underline{\Gamma_L^\uparrow} & 0 & -\Gamma_R^\uparrow & 0 & 0 & 0 & 0 & 0 & 0 & 0 & 0 & 0 \\
0 & 0 & 0 & \underline{\Gamma_L^\uparrow} & 0 & -\Gamma_R^\downarrow & 0 & 0 & 0 & 0 & 0 & 0 & 0 & -2J \\
0 & 0 & \underline{\Gamma_L^\downarrow} & 0 & 0 & 0 & -\Gamma_R^\uparrow  & 0 & 0 & 0 & 0 & 0 & 0 & 2J \\
0 & 0 & 0 & \underline{\Gamma_L^\downarrow} & 0 & 0 & 0 & -\Gamma_R^\downarrow & 0 & 0 & 0 & 0 & 0 & 0 \\
0 & 0 & 0 & 0 & 0 & 0  & 0 & 0 & -\Gamma_L & \gamma \Delta B & 0 & 0 & 0 & 0 \\
t_{12} & 0 & -t_{12} & 0 & 0 & 0 & 0 & 0 & -\gamma \Delta B & -\Gamma_L & 0 & 0 & 0 & 0 \\
0 & 0 & 0 & 0 & 0 & 0  & 0 & 0 & 0 & 0 & -\Gamma_L & -\gamma \Delta B & 0 & 0 \\
0 & t_{12} & 0 & -t_{12} & 0 & 0 & 0 & 0 & 0  & 0 & \gamma \Delta B & -\Gamma_L & 0 & 0 \\
0 & 0 & 0 & 0 & 0 & 0 & 0 & 0 & 0  & 0 & 0 & 0 & -\Gamma_R & 2 \gamma \Delta B \\
0 & 0 & 0 & 0 & 0 & J & -J & 0 & 0 & 0 & 0 & 0 & -2 \gamma \Delta B & -\Gamma_R
\end{pmatrix}.
\end{equation}
\end{widetext}
All parameters which are not included in the Liouvillian do not influence the dynamics, as long as conditions defined in the first paragraph of Sec.~\ref{sec:methods} are met. Elements included in the jump operator $\mathcal{J}_L$ are underlined and elements included in the operator $\mathcal{J}_R$ are boldfaced.

In the limit  $|t_{12}| \gg \Gamma_L, \Gamma_R, |J|, \gamma |\Delta B|$ the oscillations between the states $|0 \sigma \rangle$ and $|\sigma 0 \rangle$ are much faster then the other timescales of the system, and in result one can assume $\rho_{11}=\rho_{33}$, $\rho_{22}=\rho_{44}$. In such a case state of the system can be described using the reduced state vector $\rho_{red}=(2\rho_{11},2 \rho_{22}, \rho_{55}, \rho_{66}, \rho_{77}, \rho_{88}, \mathcal{R}_{67}, \mathcal{I}_{67})^T$. Here, two first elements of the vector correspond to the populations of the subspaces ${|\! \uparrow \rangle} \equiv \{{|\! \uparrow 0\rangle}, {|0 \! \uparrow \rangle} \}$ and ${|\! \downarrow \rangle} \equiv \{ {|\! \downarrow 0\rangle}, {|0 \! \downarrow \rangle} \}$. Dynamics of the reduced vector is generated by the reduced Liouvillian

\begin{align} \label{liovred} &\mathcal{L}_{red} = \\ \nonumber
	&\begin{pmatrix}
	-\Gamma_L & 0 & \mathbf{\Gamma_R^\uparrow} & \mathbf{\Gamma_R^\downarrow} & 0 & 0 & 0 & 0 \\
	0 & -\Gamma_L & 0 & 0 & \mathbf{\Gamma_R^\uparrow} & \mathbf{\Gamma_R^\downarrow} & 0 & 0 \\
	\underline{\Gamma_L^\uparrow/2} & 0 & -\Gamma_R^\uparrow & 0 & 0 & 0 & 0 & 0  \\
	0 & \underline{\Gamma_L^\uparrow/2} & 0 & -\Gamma_R^\downarrow & 0 & 0 & 0 & -2J \\
   \underline{ \Gamma_L^\downarrow/2} & 0 & 0 & 0 & -\Gamma_R^\uparrow & 0 & 0 & 2J \\
	0 &  \underline{\Gamma_L^\downarrow/2} & 0 & 0 & 0 & -\Gamma_R^\downarrow & 0 & 0   \\
	0 & 0 & 0 & 0 & 0 & 0 & -\Gamma_R  & 2 \delta \\
	0 & 0 & 0 & J & -J & 0 & -2 \delta & -\Gamma_R 
	\end{pmatrix},
\end{align}
where $\delta=\gamma \Delta B$. For $\Delta B=0$ it corresponds to the graphical model presented in Fig.~\ref{fig:6statescoh}~(a).

Finally, I present the Liouvillian obtained using the DME approach. Here also the limit $|t_{12}| \gg \Gamma_L, \Gamma_R, |J|, \gamma |\Delta B|$ is taken and the subspaces ${|\! \uparrow \rangle}$ and ${|\! \downarrow \rangle}$ are considered instead of the single-electron states. The state vector is defined as $\rho_{DME}=(\rho_{UU}, \rho_{DD}, \rho_{55}, \rho_{PP}, \rho_{AA}, \rho_{88})^T$, where $\rho_{UU}$ and $\rho_{DD}$ are populations of the subspaces ${|\! \uparrow \rangle}$ and ${|\! \downarrow \rangle}$, while $\rho_{PP}$ and $\rho_{AA}$ are populations of the states $\psi_P$ and $\psi_A$ defined in Eqs.~\eqref{psi1}-\eqref{psi2}. The corresponding Liouvillian reads

\begin{align} \label{liovdme} &\mathcal{L}_{DME} = \\ \nonumber
&\begin{pmatrix}
-\Gamma_L & 0 & \mathbf{\Gamma_R^\uparrow} & \mathbf{\Gamma_R^\downarrow B^{(+)}_2} & \mathbf{\Gamma_R^\downarrow B^{(-)}_2} & 0  \\
0 & -\Gamma_L & 0 & \mathbf{\Gamma_R^\uparrow B^{(-)}_2} & \mathbf{\Gamma_R^\uparrow B^{(+)}_2} & \mathbf{\Gamma_R^\downarrow}  \\
\underline{\Gamma_L^\uparrow/2} & 0 & -\Gamma_R^\uparrow & 0 & 0 & 0    \\
\underline{\Gamma_L^\downarrow B^{(-)}_4} & \underline{\Gamma_L^\uparrow B^{(+)}_4} & 0 & -\Gamma_R C^{(-)} & 0 & 0  \\
\underline{\Gamma_L^\downarrow B^{(+)}_4} & \underline{\Gamma_L^\uparrow B^{(-)}_4} & 0 & 0 & -\Gamma_R  C^{(+)} & 0   \\
0 &  \underline{\Gamma_L^\downarrow/2} & 0 & 0 & 0 & -\Gamma_R^\downarrow    \\
\end{pmatrix},
\end{align}
where $B_2^{(\pm)}=(1 \pm \sin \theta)/2$, $B_4^{(\pm)}=(1 \pm \sin \theta)/4$ and $C^{(\pm)}=1 \pm p_R \sin \theta$. It corresponds to the graphical model of the dynamics presented in Fig.~\ref{fig:6stanowdme}.

Among the Liouvillians presented above, this given by Eq.~\eqref{fullliov} is the most general and captures all physics of transport in the regime considered in this paper. The Liouvillian given by Eq.~\eqref{liovred} is valid in the limit $|t_{12}| \gg \Gamma_L, \Gamma_R, |J|, \gamma |\Delta B|$. It simplifies the qualitative discussion of the dependence of the studied quantities on $J$. The Liouvillian given by Eq.~\eqref{liovdme} is the least general -- it works well only when the condition $|t_{12}| \gg \sqrt{J^2+(\gamma \Delta B)^2} \gg \Gamma_R$ is met. However, it simplifies the qualitative discussion of the dependence of the dynamics on the angle $\theta$.

\end{document}